\documentclass[12pt,draft]{article}
\usepackage{graphicx}
\usepackage{amsmath}
\usepackage{a4}

\input{tcilatex}

\begin{document}

\begin{center}
\textbf{THE PARTITION FUNCTION FOR ''COMPOSITE\ PARTICLES''}

\bigskip

by

\bigskip

M.\ C.\ Berg\`{e}re

\bigskip

Service de Physique Th\'{e}orique, CEA-Saclay

F-91191 Gif sur Yvette cedex, France

\bigskip email: bergere@spht.saclay.cea.fr

\bigskip

\bigskip
\end{center}

\textbf{Abstract: }We calculate the partition function for ''composite
particles''. For any finite number of states $d,$ and in the following two
cases: 1$%
{{}^\circ}%
)$ all states have the same energy, 2$%
{{}^\circ}%
)$ the energy is linearly distributed over the states, we transform the
partition function into a finite sum of terms which exhibit trivially their $%
d\;$dependance. The infinite $d$ thermodynamic limit is obtained as well as
the finite $d$-size corrections, etc... In the second case, we obtain the
finite $d$-size corrections to ''the universal chiral partition function for
exclusion statistics''.

\section{\protect\bigskip Introduction}

Fractional statistic was introduced in the literature in order to explain
various effects encountered in the physics of condensed matter such as the
fractional quantum Hall effect, the high temperature superconductivity, the
properties of the spin 1/2 antiferromagnetic chains...and was explicitely
observed in several theoretical models like for instance the two-dimensional
anyon model$^{\left[ 1\right] }$ or the one-dimensional Calogero-Sutherland
model$^{\left[ 2\right] }.$ This statistic generalizes the exclusion Pauli
principle and interpolates between bosons and fermions statistics in the
sense that it takes $g$ states to add one more particle to the system ($g=0$
for bosons and$\;g=1$ for fermions) where $g$ is fractional $(g=l/m);$
Haldane$^{\left[ 3\right] }$ proposed that in the fractional statistic, the
number of possible configurations for $N\;$particles on $d$ states should be 
\begin{equation}
C_{d+\left( 1-g\right) \left( N-1\right) }^{N}  \label{1}
\end{equation}
Of course, this formula makes sense only statistically for large $N$ and $d$%
, otherwise $\left[ d+\left( 1-g\right) \left( N-1\right) \right] $ is a
fraction and the combinatoric $C$ symbol calculated with the $\Gamma $ Euler
function gives a fractional number of configurations. However, we note that
for finite $N$ of the form $N=pm+1$, the combinatoric $C$ symbol makes sense
and gives an integer number of configurations. This remark leads us to
propose$^{\left[ 4\right] }$ a specific definition of the ''composite
particles'' which naturally generates the fractional statistic:

A ''composite particle'' is a set of $m$ particles and $d\geq l$ states with
the constraint that the $\left( l-1\right) $ bottom states are empty and the 
$l^{th}$ state (from the bottom) has at least one particle. We define an
uncomplete ''composite particle'' as a set of $0<r\leq m$ \ particles with
no constraints on the empty states.

The main constraint in this construction is that a non empty state is
entirely included inside a ''composite particle'' and cannot be splitted
over several ''composite particles'' (fig.1). The number of possible
configurations for the above definition of ''composite particles'' has been
found in $^{\left[ 4\right] }$ as 
\begin{equation}
C_{d+N-1-lE\left( \frac{N-1}{m}\right) }^{N}  \label{2}
\end{equation}
where $E\left( x\right) $ is the integer part of $x$. If $N=pm+1$, this
formula gives the same result as Haldane's formula, but (2) makes sense for
any $N=pm+r\;(0<r\leq m)$. For large $N$, the value of $r$ is irrelevant and
the asymptotic properties of (1) and (2) are equivalent.\ In that sense, the
''composite particles'' reproduce the fractional statistic introduced by
Haldane and extend it naturally to any finite $N$ and $d$; in addition, it
gives a geometrical interpretation of this statistic.

If all states have the same energy $E,$ it is found that for large $N$ and $%
d $ $(\;\lg \left( N!\right) \sim N\lg \left( N/e\right) $ $),$ the average
number of particles per state is given by $^{\left[ 5\right] }$%
\begin{equation}
n\left( y,g\right) \;=\frac{1}{W\left( y,g\right) +g}<\frac{1}{g}  \label{3}
\end{equation}
where $y=\exp \left( -E/kT\right) $ and where the function $W\left(
y,g\right) $ satisfies 
\begin{equation}
\left[ W\left( y,g\right) \right] ^{g}\;\left[ 1+W\left( y,g\right) \right]
^{^{1-g}}=y^{-1}  \label{4}
\end{equation}
In appendix A, we show that 
\begin{equation}
W\left( y,g\right) =y^{-1}+g-1+\left( g-1\right) \sum_{n=1}^{\infty }\frac{%
ng\left( ng-1\right) ...\left( ng-n+1\right) }{\left( n+1\right) !}\;\left(
-y\right) ^{n}  \label{5}
\end{equation}
as already found by Ramanujan$^{\left[ 6\right] }.$

\bigskip

More recently$^{\left[ 7\right] }$, it has been proved that the ''composite
particles'' with $N$ particles were in a one-to-one correspondance with the
Young tableaux with at most $N$ rows. As a consequence, and using the Bethe
Ansatz, it was possible to reproduce from the ''composite particles'', the
eigenstates of the Hamiltonians for the models of Calogero-Sutherland$^{%
\left[ 2\right] }$ and of Ruijsenaars-Schneider$^{\left[ 8\right] }$. From
the complete knowledge of the spectral properties of these models, it was
then possible to define a momentum space representation for the ''composite
particles'' in terms of creation operators attached to some Young tableaux;
these creation operators are generated from the normal ordered product of
vertex operators when expanded in terms of Jack or Macdonald$^{\left[ 9%
\right] }$ homogeneous, symmetric polynomials. Similarly, a position space
representation has been proposed for the ''composite particles'' in terms of
appropriate products of vertex operators.

\bigskip

In this publication, we calculate the partition function for the ''composite
particles'' and for any finite number of states $d,\;$in the following two
cases:

1$%
{{}^\circ}%
)$ when all the states have the same energy $E,$ we find 
\begin{equation}
Z\left( y,l,m\right) =\sum_{j=0}^{Sup\left( m,l\right) -1}\alpha _{j}\left(
y,l,m\right) \;\;\left[ \beta _{j}\left( y,l,m\right) \right] ^{d}  \label{6}
\end{equation}
where $y=\exp \left( -E/kT\right) .$ We determine the $d$-independant
functions $\alpha _{j}\left( y,l,m\right) $ and $\;\beta _{j}\left(
y,l,m\right) ;$ this result generalizes the bosons partition function $%
\left( 1-y\right) ^{-d},$ and the fermions partition function $\left(
1+y\right) ^{d}.$

$\;$2$%
{{}^\circ}%
)$ when the energy is linearly distributed over the states, that is when the
state $i\ $contributes to the partition function by a factor 
\begin{equation}
x^{i-1}y=\exp \left[ -\left( E+\left( i-1\right) v\right) /kT\right]
\;\;\;\;v\geq 0  \label{7}
\end{equation}
In that case, we find 
\begin{equation}
Z\left( x,y,l,m\right) =\sum_{j=0}^{Sup\left( l,m\right) -1}\alpha
_{j}\left( y,x,l,m\right) \;\left[ \prod_{i=0}^{d-1}\beta _{j}\left(
x^{i}y,x,l,m\right) \right] \;\gamma _{j}\left( x^{d}y,x,l,m\right)
\label{8}
\end{equation}
and we determine the $d$ independant functions $\alpha _{j}(y,x,l,m),$ $%
\beta _{j}\left( y,x,l,m\right) $ and $\gamma _{j}\left( y,x,l,m\right) .$\
The case 2$%
{{}^\circ}%
)$ is essentially a $x$-deformation of the case 1$%
{{}^\circ}%
).\;$In both cases, we transform the partition function for any finite $d,$
into a finite sum of terms which exhibits trivially their $d$ dependance;
consequently, our result provides naturally the large $d$ asymptotic
behaviour, the finite $d-$size corrections, etc...of the partition function.

\bigskip

In the litterature, the function $\beta _{0}\left( y,l,m\right) $ in (6)
which describes the large $d$ thermodynamic limit is known$^{\left[ 5\right]
}$ as 
\begin{equation}
\beta _{0}\left( y,l,m\right) =1+V\left( y,l/m\right)  \label{9}
\end{equation}
where $V\left( y,l/m\right) =\left[ W\left( y,g\right) \right] ^{-1}$
satisfies the equation 
\begin{equation}
V\left( 1+V\right) ^{g-1}=y  \label{10}
\end{equation}
On the other hand, we define the thermodynamic limit in case 2$%
{{}^\circ}%
)$ as the limit $x=\xi ^{\frac{1}{d}}\rightarrow 1$ and $d\rightarrow \infty
;$ we obtain for the leading contribution 
\begin{equation}
\lim d\rightarrow \infty \;\prod_{i=0}^{d-1}\beta _{0}\left( \xi ^{\frac{i}{d%
}}y,\xi ^{\frac{1}{d}},l,m\right) \sim \exp \left( \frac{d}{\ln \xi ^{-1}}%
\left[ B\left( y\right) -B\left( \xi y\right) \right] \right)  \label{11}
\end{equation}
where 
\begin{eqnarray}
B\left( y\right) &=&\int_{0}^{y}\frac{dv}{v}\ln \left[ 1+V\left( v,g\right) %
\right]  \notag \\
B\left( y\right) &=&L\left[ \frac{V\left( y,g\right) }{1+V\left( y,g\right) }%
\right] +\frac{1}{2}\ln y\;.\;\ln \left[ 1+V\left( y,g\right) \right]
\label{12}
\end{eqnarray}
and where $L\left( y\right) $ is the Rogers dilogarithm function. The Rogers
dilogarithm function and dilogarithm identities were previously used by
Nahm, Recknagel and Terhoeven$^{\left[ 10\right] }$ to calculate the
structure constants and, in particular, the effective central charge of
certain conformal field theories. Later, Hikami$^{\left[ 11\right] }$
calculated the effective central charge for an ideal $g$-on gas as 
\begin{equation}
c=\frac{6}{\pi ^{2}}\int_{0}^{1}\frac{dv}{v}\ln \left[ 1+V\left( v,g\right) %
\right] =\frac{L\left( \xi _{H}^{g}\right) }{L\left( 1\right) }  \label{13}
\end{equation}
where $\xi _{H}$ satisfies the equation 
\begin{equation}
\xi _{H}^{g}=1-\xi _{H}  \label{14}
\end{equation}
More recently, the function $B\left( y\right) $ was obtained by Berkovich
and McCoy$^{\left[ 12\right] }$ using the steepest descent method in order
to evaluate the thermodynamic limit of what they call ''the universal chiral
partition function for exclusion statistics''.Their partition function was
defined in the context of conformal field theory and of Rogers-Ramanujan
identities. They demonstrate the equality between their partition function
and Haldane's exclusion statistic. ''The universal chiral partition function
for exclusion statistics'' has been defined for several different kinds of
particles where $g$ is replaced by a matrix $G_{ab}$; in Ref $\left[ 13%
\right] $ several authors have explored the properties of these statistics
in order to apply them to physical situations in condensed matter physics.

\bigskip

In section 2, we calculate the partition function when all the states have
the same energy. In section 3, we give the general expression of the
partition function, when the state $i$ contributes to a factor $x_{i}=\exp
\left( -E_{i}/kT\right) ,$ as a sum of non-symmetric polynomials of the $%
x_{i}$'s. In section 4, we apply section 3 to the case where the energy is
linearly distributed over the states as in equation (7); in that case, the
partition function can be written as the following sum 
\begin{equation}
Z\left( x,y,l,m\right) =1+\sum_{p=0}^{E\left( \frac{d-1}{l}\right)
}\sum_{r=1}^{m}\;x^{\left( \frac{p\left( p-1\right) }{2}m+pr\right)
l}\;C_{d+p\left( m-l\right) +r-1}^{pm+r}\left( x\right) \;\;\;y^{pm+r}
\label{15}
\end{equation}
It must be noted at that point that ''the universal chiral partition
function for exclusion statistics'' as considered by Berkovich and McCoy$^{%
\left[ 12\right] }$ and which originated in the work of Kedem, Klassen, Mc
Coy and Melzer$^{\left[ 12\right] }$ in the context of conformal field
theory, does contain the symbol $C_{d+\left( 1-g\right) \left( N-1\right)
}^{N}\left( x\right) $ (generalized to several kind of particles with a
matrix $g$); however, their calculation of the thermodynamic limit is made
for one kind of particles at infinite$\;d$ so that the symbol $C_{d+\left(
1-g\right) \left( N-1\right) }^{N}\left( x\right) $ is replaced by $\left[
\left( x,x\right) _{N}\right] ^{-1}$ where $\left( x,x\right)
_{n}=\prod_{i=1}^{n}\left( 1-x^{i}\right) $ and their limit when $%
x\rightarrow 1$ is a special case of ours.

In the rest of section 4, we transform (15) into (8) for any finite number $%
d $ of states.

\bigskip

\bigskip

Let us conclude this introduction by mentionning the fact that our
construction of the ''composite particles'', for any finite number of
particles $N$ and for any finite number of states $d,\;$reproduces perfectly
well previous results known in the literature, provides finite $d$-size
corrections to these results and, maybe more important, gives a quasi
geometrical interpretation to fractional statistic.

\bigskip

\bigskip

$\ \ \ \ \ \ \ \ \ \ \ \ \ \ \ \ \ \ \ \ \ \ \ \ \ \ $

\section{The partition function Z$\left( y,l,m\right) $}

The partition function 
\begin{equation}
Z\left( y,l,m\right) =\sum_{N=0}^{mE\left( \frac{d+l-1}{l}\right)
}C_{d+N-1-lE\left( \frac{N-1}{m}\right) }^{N}\;\;\;y^{N}  \label{16}
\end{equation}
may also be written in terms of the number of ''composite particles'' p and
of the remaining number of particles r in the uncomplete ''composite
particle'': 
\begin{equation}
Z\left( y,l,m\right) =1+\sum_{p=0}^{E\left( \frac{d-1}{l}\right)
}\sum_{r=1}^{m}\;C_{d+p\left( m-l\right) +r-1}^{pm+r}\;\;\;y^{pm+r}
\label{17}
\end{equation}
The sums in (17) can be calculated for any finite number of states $d$. The
purpose of this section is to transform the polynomial in $y$ which defines
the partition function into a finite expansion of the type 
\begin{equation}
Z\left( y,l,m\right) =\sum_{j=0}^{Sup\left( l,m\right) -1}\alpha _{j}\left(
y,l,m\right) \;\left[ \beta _{j}\left( y,l,m\right) \right] ^{d}\;
\label{18}
\end{equation}
where the sum over $j$ is finite and where the functions $\alpha _{j}\left(
y,l,m\right) $ and $\beta _{j}\left( y,l,m\right) $ are $d$ independant. The
easiest examples are given by the cases of the bosons and the fermions
statistics where 
\begin{equation}
Z_{B}=\sum_{N=0}^{\infty }\;C_{d+N-1}^{N}\;y^{N}=\frac{1}{\left( 1-y\right)
^{d}}  \label{19}
\end{equation}
\begin{equation}
Z_{F}=\sum_{N=0}^{d}\;C_{d}^{N}\;y^{N}=\left( 1+y\right) ^{d}  \label{20}
\end{equation}
In the general case, the structure of the function $Z(y,l,m)\;$is twofold. A
first effect can be traced in the special case $m=1$ and results in the
introduction of functions related to $W(y,g)$ defined in (5). A second
effect appears when $m>1$ and is expressable as a sum of similar expressions
which differ by the choice of a $m^{th}$ root of unity. This second effect
becomes particularly simple for $l=m$ since in that case the function $%
W(y,g=1)=y^{-1}.$

Let us first consider the second effect that is the case of ''composite
fermions'' where $l=m>0\;$ or $g=1.$ We define 
\begin{equation}
Z_{r}\left( y,m,m\right) =\delta _{r,m}+\sum_{p=0}^{E\left( \frac{d-1}{m}%
\right) }\;C_{d+r-1}^{pm+r}\;\;\;y^{pm+r}  \label{21}
\end{equation}
An easy calculation gives 
\begin{equation}
Z_{r}\left( y,m,m\right) =\frac{1}{m}\sum_{n=0}^{m-1}q^{-rn}\left(
1+q^{n}y\right) ^{d+r-1}  \label{22}
\end{equation}
where $q=\exp \left( 2i\pi /m\right) .$ Finally, summing over $r$ from $1$
to $m$, we get the partition function for the ''composite fermions'' as 
\begin{equation}
Z\left( y,m,m\right) =\frac{1}{m}\sum_{n=0}^{m-1}\frac{\left(
1+q^{n}y\right) ^{m}-1}{1+q^{n}\left( y-1\right) }\;\left( 1+q^{n}y\right)
^{d}  \label{23}
\end{equation}
which is of the form (18).

The thermodynamic limit, at large $d$ is dominated by the root of unity
equal to $1$ that is $n=0$; we get 
\begin{equation}
Z_{th}\left( y,m,m\right) =\frac{1}{m}\frac{\left( 1+y\right) ^{m}-1}{y}%
\;\left( 1+y\right) ^{d}  \label{24}
\end{equation}
which shows that we recover the usual fermion partition function corrected
by $m$ dependant ''finite $d-$size effects''. At zero energy, this
correcting factor is simply $(2^{m}-1)/m.$

\bigskip

The first effect (for $m=1$) is obtained when calculating 
\begin{equation}
Z\left( y,l,1\right) =\sum_{N=0}^{E\left( \frac{d+l-1}{l}\right)
}\;C_{d+\left( 1-l\right) \left( N-1\right) }^{N}\;\ \;y^{N}  \label{25}
\end{equation}
This sum can be calculated either from a recurrence on $Z$ as a function of $%
d$ (like in section 4.3 and appendix B) or by introducing a contour integral
representation of the combinatoric symbols as already presented in Ref. $%
\left[ 14\right] $%
\begin{equation}
C_{n}^{p}=\frac{1}{2i\pi }\oint_{C_{0}}dz\;\frac{\left( 1+z\right) ^{n}}{%
z^{p+1}}  \label{26}
\end{equation}
where the contour $C_{0}\;$surrounds the point $z=0.$ It is useful to note
that if the contour surrounds $z=0$ and $z=-1,$ then the contour integral
gives zero for $N>E\left( \frac{d+l-1}{l}\right) $ so that the geometrical
series in $N$ can be summed to infinity provided that the series converge on
the contour of integration. We obtain for $l>0$%
\begin{equation}
Z\left( y,l,1\right) =\frac{1}{2i\pi }\oint_{C_{l}}dz\;\frac{\left(
1+z\right) ^{d+2l-2}}{z\left( 1+z\right) ^{l-1}-y}\;\;\;l>0\;\;  \label{27}
\end{equation}
The convergence of the series requires $\left| z\right| \left| 1+z\right|
^{l-1}>y$ \ which is the case for large $\left| z\right| $\ if $l>0$ (we
remind that $0<y\leq 1$).\ Consequently, the contour $C_{l}$ surrounds the $%
l\;$roots $V_{j}\left( y,l\right) \;$of the polynomial equation 
\begin{equation}
z\left( 1+z\right) ^{l-1}=y  \label{28}
\end{equation}
If we expand the denominator in (27) around each pole $z=V_{j}\left(
y,l\right) $, we obtain 
\begin{equation}
Z\left( y,l,1\right) =\sum_{j=0}^{l-1}\frac{\left[ 1+V_{j}\left( y,l\right) %
\right] ^{d+l}}{1+l\;V_{j}\left( y,l\right) }\;\;\;l>0  \label{29}
\end{equation}
which is again of the form (18).

Several remarks should be made at this stage:

1$%
{{}^\circ}%
)$ the $l$ roots $V_{j}\left( y,l\right) $ are independant of the number of
states $d,$

2$%
{{}^\circ}%
)$ when $y\rightarrow 0,$ the root $V_{0}\left( y,l\right) \sim y;$ the $%
\left( l-1\right) $ other roots $\rightarrow -1+O\left( y^{\frac{1}{l-1}%
}\right) .$ All the roots can be expressed in terms of a unique function $%
V\left( y,l\right) $ defined in appendix A:

\begin{eqnarray}
V_{0}\left( y,l\right) &=&V\left( y,l\right)  \notag \\
V_{j}\left( y,l\right) &=&-1-V\left( Q^{2j+l}\;y^{\frac{1}{l-1}},\frac{l}{l-1%
}\right) \;\;\;\;\;\;\;j=1,...,l-1  \label{30}
\end{eqnarray}
where $Q=\exp \left( i\pi /\left( l-1\right) \right) ,$

3$%
{{}^\circ}%
)$ the sum over all roots is such that $Z\left( y,l,1\right) $ is a
polynomial in $y$ of degree $E\left( \frac{d+l-1}{l}\right) ;$ the roots $%
V_{j}\left( y,l\right) $ for $j>0$ contribute to powers of $y>E\left( \frac{%
d+l-1}{l}\right) ,$

4$%
{{}^\circ}%
)$ from equation (28) and the zero $y\;$limit, we see that $V\left(
y,l\right) =\left[ W\left( y,l\right) \right] ^{-1}$ defined in (5).

\bigskip

\bigskip

In the thermodynamic large $d$ limit, the polynomial in y which defines the
partition function in (25), becomes an infinite series which can be summed
into 
\begin{equation}
Z_{th}\left( y,l,1\right) =\frac{\left[ 1+V\left( y,l\right) \right] ^{d+l}}{%
1+l\;V\left( y,l\right) }\;\;\;l\geq 0  \label{31}
\end{equation}
expressed in terms of a unique function $V\left( y,\alpha \right) $ defined
in appendix A:This result is also valid for $l=0\;$where $V\left( y,l\right)
=y/\left( 1-y\right) .$

\bigskip

\bigskip

We now write the expression for the partition function $Z\left( y,l,m\right)
\;$defined by the sum (17) for any $l$ and $m.$ This partition function
contains both effects described above.\ The details of the proof are given
in appendix A; we get 
\begin{equation}
Z\left( y,l,m\right) =\sum_{j=0}^{Sup\left( m,l\right) -1}\;\frac{\left(
1+V_{j}\right) ^{l}-1}{\left( 1+V_{j}\right) ^{\frac{l}{m}}-q^{k\left(
j\right) }}\;\frac{\left( 1+V_{j}\right) ^{d+\frac{l}{m}}}{m+l\;V_{j}}
\label{32}
\end{equation}
where $q=\exp \left( 2i\pi /m\right) $ and where $k\left( j\right) $ is an
integer equal to $j$ for $j=0,...,m-1$ but depends upon the definition of $%
(1+V_{j})^{1/m}$ in (A17) for $l>m$ and $j=m,...,l-1.$ The functions $V_{j}$
are the roots of the polynomial equations 
\begin{equation}
z^{m}-\left( 1+z\right) ^{m-l}\;y^{m}=0\;\;\;\;\;m\geq l  \label{33}
\end{equation}
\begin{equation}
z^{m}\left( 1+z\right) ^{l-m}-y^{m}=0\;\;\;\;\;l\geq m\;  \label{34}
\end{equation}
and may be written in terms of a unique function $V\left( y,\alpha \right) $
defined in appendix A.

\bigskip

In the thermodynamic large $d$ limit, the partition function is dominated
(at least for $y$ small enough) by the root $V_{0,}$ so that 
\begin{equation}
Z_{th}\left( y,l,m\right) =\frac{\left( 1+V\left( y,l/m\right) \right) ^{l}-1%
}{\left( 1+V\left( y,l/m\right) \right) ^{\frac{l}{m}}-1}\;\frac{\left(
1+V\left( y,l/m\right) \right) ^{d+\frac{l}{m}}}{m+l\;V\left( y,l/m\right) }
\label{35}
\end{equation}

The average number of particles $\overline{N}$ is easily obtained from 
\begin{equation}
\overline{N}=\frac{1}{Z}\;y\frac{\partial Z}{\partial y}  \label{36}
\end{equation}
and 
\begin{equation}
y\frac{\partial V\left( y,l/m\right) }{\partial y}=\frac{V\left(
y,l/m\right) \left( 1+V\left( y,l/m\right) \right) }{1+\frac{l}{m}\;V\left(
y,l/m\right) }  \label{37}
\end{equation}
In the large $d$ limit, we obtain the average number of particles per state
as 
\begin{equation}
n\left( y,g\right) =\lim \;d\rightarrow \infty \;\frac{\overline{N}}{d}=%
\frac{\;V\left( y,g\right) }{1+g\;V\left( y,g\right) }  \label{38}
\end{equation}
which shows again from (3) that $V\left( y,g\right) =\left[ W\left(
y,g\right) \right] ^{-1}.$

\section{\textbf{The general structure of the partition function}}

\bigskip

We now construct the partition function when the state $i$ has a given
energy $E_{i}$ for $i=1,...,d.$ We introduce the reduced variables $%
x_{i}=\exp \left( -E_{i}/kT\right) .$\ We first consider the uncomplete
''composite particle'' with $r$ particles on $\delta $ states. Its
contribution to the partition function is 
\begin{equation}
y^{r}\sum_{\left\{ r_{i\geq 0}\right\} }\;x_{1}^{r_{1}}...x_{\delta
}^{r_{\delta }}  \label{39}
\end{equation}
where $r_{i}$ is the occupation number of the state $i$ and $y$ is a
variable which counts the number of particles. Of course the sum in (39) is
restricted to $\sum_{i=1}^{\delta }r_{i}=r.$ This contribution can be
written in terms of the symmetric, homogeneous (of degree $r$) Schur
polynomial attached to the Young tableau with one row and $r$ squares

\begin{equation}
y^{r}\;S_{r}\left( x_{1},...,x_{\delta }\right)  \label{40}
\end{equation}
Now, a ''composite particle'' with $\delta $ states has $m$ particles on $%
\delta -l+1$ states; moreover, we have to make sure that the state ''$\delta
-l+1$'' has at least one particle; consequently, the contribution of such a
''composite particle'' to the partition function is 
\begin{equation}
y^{m}\left[ S_{m}\left( x_{1},...,x_{\delta -l+1}\right) -S_{m}\left(
x_{1},...,x_{\delta -l}\right) \right] =y^{m}\;x_{\delta
-l+1}\;S_{m-1}\left( x_{1},...,x_{\delta -l+1}\right)  \label{41}
\end{equation}
We now organize the $p$ ''composite particles'' each of which having $\delta
_{i}$ states ($i=1,...,p)$ and the uncomplete ''composite particle'' with $%
\delta _{p+1}$ states. Let us introduce the notation 
\begin{equation}
\Delta _{k}=\sum_{n=1}^{k}\;\delta _{n}  \label{42}
\end{equation}
Then, the partition function is 
\begin{eqnarray}
Z &=&1+\sum_{p=0}^{E(\frac{d-1}{l})}\sum_{r=1}^{m}\;y^{pm+r}\sum_{\left\{
\delta _{i}\right\} }\;(\;\prod_{k=1}^{p}\left( x_{\Delta
_{k}-l+1}\;S_{m-1}\left( x_{\Delta _{k-1}+1},...,x_{\Delta _{k}-l+1}\right)
\right) \;.  \notag \\
&&\;.\;\;S_{r}\left( x_{\Delta _{p}+1},...,x_{d}\right) )  \label{43}
\end{eqnarray}
where the sum over $\left\{ \delta _{i}\right\} $ is restricted by $\Delta
_{p+1}=d$ \ where $d$ is the total number of states. The sum over $\left\{
\delta _{i}\right\} $ may be performed to obtain 
\begin{equation}
Z=1+\sum_{p=0}^{E\left( \frac{d-1}{l}\right) }\sum_{r=1}^{m}\;Z_{N}\left(
x_{1},...,x_{d}\right) \;\;y^{pm+r}  \label{44}
\end{equation}
where $Z_{N}\left( x_{1},...,x_{d}\right) $ is a homogeneous (of degree $%
N=pm+r$), non symmetric polynomial of the variables $x_{i}:$%
\begin{equation}
Z_{N}\left( x_{1},...,x_{d}\right) =\sum_{\left\{ p_{i}\right\} \in \Lambda
_{N,m,l}}x_{1}^{p_{1}}...x_{d}^{p_{d}}  \label{45}
\end{equation}
In (45), the set $\Lambda _{N,m,l}$ of possible $p_{i}$'s is defined by
three constraints:

1$%
{{}^\circ}%
)$ 
\begin{equation}
\sum_{i=1}^{d}\;p_{i}=N  \label{46}
\end{equation}

2$%
{{}^\circ}%
)$ let 
\begin{equation}
\pi _{i}=\sum_{j\leq i}\;p_{j}\;\;\;\;i=1,...,d  \label{47}
\end{equation}
then, 
\begin{equation}
\left\{ m,2m,...,pm\right\} \subseteq \left\{ \pi _{i}\right\}  \label{48}
\end{equation}

\bigskip

3$%
{{}^\circ}%
)$ we consider the indices $\left( i_{1},...,i_{p}\right) $ such that $\pi
_{i_{q}}=qm,$ then

\begin{equation}
p_{i_{q}+1}=p_{i_{q}+2}=...=p_{i_{q}+l-1}=0\;\;\;\;\;\;\;q=1,...,p
\label{49}
\end{equation}

The number of monomials in $Z_{N}$ is $C_{d+N-1-lE\left( \frac{N-1}{m}%
\right) }^{N}.$

\bigskip

\section{\textbf{The partition function when the energy is linearly
distributed over the states.}}

\bigskip

In this section, we apply the structure obtained in section 3 to the case
where the energy for the state $i$ is 
\begin{equation}
E_{i}=E+\left( i-1\right) v\;\;\;\;\;v\geq 0,\;\;\;i=1,...,d  \label{50}
\end{equation}
The reduced variables are defined as $x^{i-1}y=\exp \left( -E_{i}/kT\right)
. $ The Schur polynomials in (43)\bigskip\ are known to be 
\begin{equation}
S_{n}\left( 1,x,x^{2},...,x^{\delta -1}\right) =C_{\delta +n-1}^{n}\left(
x\right)  \label{51}
\end{equation}
where 
\begin{equation}
C_{n}^{p}\left( x\right) =\frac{\left( x;x\right) _{n}}{\left( x;x\right)
_{p}\left( x;x\right) _{n-p}}  \label{52}
\end{equation}
and 
\begin{equation}
\left( a;x\right) _{n}=\left( 1-a\right) \left( 1-ax\right) ...\left(
1-ax^{n-1}\right)  \label{53}
\end{equation}
Using the homogeneity property of the Schur polynomials, we obtain for the
partition function 
\begin{equation}
Z\left( x,y,l,m\right) =1+\sum_{p=0}^{E\left( \frac{d-1}{l}\right)
}\sum_{r=1}^{m}y^{pm+r}\sum_{\left\{ \delta _{i}\right\} }x^{\nu \left(
\delta _{i}\right) }\prod_{i=1}^{p}C_{\delta _{i}+m-l-1}^{m-1}\left(
x\right) \;.\;\;C_{\delta _{p+1}+r-1}^{r}\left( x\right)  \label{54}
\end{equation}
where 
\begin{equation}
\nu \left( \delta _{i}\right) =m\sum_{i=1}^{p-1}\Delta _{i}+\left(
r+1\right) \Delta _{p}-pl  \label{55}
\end{equation}
and $\Delta _{i}$ is given in (42). Next, we must sum over all number of
states ($\delta _{i}\geq $ $l$ \ if $1\leq i\leq p,$ and $\delta _{p+1}\geq $
$1)$ with the restriction that $\Delta _{p+1}=d.$ To perform this sum, we
apply the usual technique of the generating functional; we multiply $%
\sum_{\left\{ \delta _{i}\right\} }...$ in (54) by 
\begin{equation}
z_{1}^{\delta _{1}-l}...z_{p}^{\delta _{p}-l}z_{p+1}^{\delta _{p+1}-1}
\label{56}
\end{equation}
we sum all $\delta _{i}$'s from $l$ to $\infty \;(i=1,...,p)$ and $\delta
_{p+1}$ from $1$ to $\infty $ and we apply the relation 
\begin{equation}
\sum_{\delta =0}^{\infty }C_{n+\delta -1}^{\delta }\left( x\right)
\;a^{\delta }=\frac{1}{\left( a;x\right) _{n}}  \label{57}
\end{equation}
We obtain for $\sum_{\left\{ \delta _{i}\right\} }...$in (54) 
\begin{equation}
x^{\left( \frac{p\left( p-1\right) }{2}m+pr\right) l}\prod_{i=1}^{p}\left( 
\frac{1}{\left( x^{\alpha _{i}}z_{i};x\right) _{m}}\right) .\frac{1}{\left(
z_{p+1};x\right) _{r+1}}  \label{58}
\end{equation}
where $\alpha _{i}=\left( p-1\right) m+r+1.$ If we set all $z_{i}$'s equal
to $z$ in (58), we get 
\begin{equation}
x^{\left( \frac{p\left( p-1\right) }{2}m+pr\right) l}\frac{1}{\left(
z;x\right) _{pm+r+1}}  \label{59}
\end{equation}
Finally, the partition function is obtained by determining the term of (59)
in $z^{d-pl-1}.$ Using (57) backwards, we get 
\begin{equation}
Z\left( x,y,l,m\right) =1+\sum_{p=0}^{E\left( \frac{d-1}{l}\right)
}\sum_{r=1}^{m}\;x^{\left( \frac{p\left( p-1\right) }{2}m+pr\right)
l}\;C_{d+p\left( m-l\right) +r-1}^{pm+r}\left( x\right) \;\;\;y^{pm+r}
\label{60}
\end{equation}
This expression is ''the universal chiral partition function'' for one kind
of particles.

In the remaining part of section 4, we compute the sums over $p$ and $r$ in
(60) for any $x,y$ and finite $d$ in the same way as in section 2. We first
remind to the reader the special cases of the bosons and of the fermions;
then, we distinguish, like in section 2, the case $l=m\;(g=1)$ and the case $%
m=1,$ before writing the general result which is proved in Appendix B.

\bigskip

\subsection{Bosons and Fermions}

\bigskip

The partition function for bosons is given by the expansion (60) with $l=0$%
\begin{equation}
Z_{B}=\sum_{N=0}^{\infty }\;C_{d+N-1}^{N}\left( x\right) \;y^{N}=\frac{1}{%
\left( y;x\right) _{d}}  \label{61}
\end{equation}
Similarly, the fermions partition function is given by (60) with $l=m=1$%
\begin{equation}
Z_{F}=\sum_{N=0}^{d}\;x^{\frac{N\left( N-1\right) }{2}}\;C_{d}^{N}\left(
x\right) \;y^{N}=\left( -y;x\right) _{d}\;  \label{62}
\end{equation}
Clearly, for $x=1,$ we recover the well known results of section 2. The
large $d$ limit for $x<1$ is given by $\left[ \left( y;x\right) _{\infty }%
\right] ^{-1}$ for bosons and $\left( -y;x\right) _{\infty }$ for fermions.
These finite results (which are not exponentionnally large with $d$)\ imply
that the average number of particles $\overline{N}$ defined in (36) is a
finite quantity, so that the average number of particles per state $n(x,y,g)$
defined in (38)$\;$is zero.\ An interesting thermodynamic limit is obtained
when $x\rightarrow 1$ and $d\rightarrow \infty .\;$Let us write $x=\xi ^{%
\frac{1}{d}}$ so that (7) becomes 
\begin{equation}
\xi ^{\frac{i-1}{d}}y=\exp \left[ -\left( E+\frac{i-1}{d}\left( E_{\infty
}-E\right) \right) /kT\right]  \label{63}
\end{equation}
At large $d,$ the energy spectrum becomes linearly continuous from $E$ to $%
E_{\infty }.$ Using the exponential representation 
\begin{equation}
\frac{1}{\left( y;x\right) _{d}}=\exp \left( \sum_{n=1}^{\infty }\;\frac{%
1-x^{dn}}{1-x^{n}}\;\frac{y^{n}}{n}\right)  \label{64}
\end{equation}

\begin{equation}
\left( -y;x\right) _{d}=\exp \left( -\sum_{n=1}^{\infty }\;\frac{1-x^{dn}}{%
1-x^{n}}\;\frac{\left( -y\right) ^{n}}{n}\right)  \label{65}
\end{equation}
and performing a $1/d$ expansion, we obtain the thermodynamic limit for the
bosons as 
\begin{equation}
Z_{B,th}=\exp \left( \frac{d}{\ln \xi ^{-1}}\sum_{n=1}^{\infty }\left( 1-\xi
^{n}\right) \;\frac{y^{n}}{n^{2}}+\frac{1}{2}\sum_{n=1}^{\infty }\left(
1-\xi ^{n}\right) \frac{y^{n}}{n}+O\left( \frac{1}{d}\right) \right)
\label{66}
\end{equation}
that is 
\begin{equation}
Z_{B,th}=\sqrt{\frac{1-\xi y}{1-y}}\exp \left( \frac{d}{\ln \xi ^{-1}}\left[
Li_{2}\left( y\right) -Li_{2}\left( \xi y\right) \right] \right) +O\left( 
\frac{1}{d}\right)  \label{67}
\end{equation}
where the dilogarithm function is 
\begin{equation}
Li_{2}\left( y\right) =-\int_{0}^{y}\frac{du}{u}\ln \left( 1-u\right)
\label{68}
\end{equation}
Similarly, for the fermions 
\begin{equation}
Z_{F,th}=\exp \left( -\frac{d}{\ln \xi ^{-1}}\sum_{n=1}^{\infty }\left(
1-\xi ^{n}\right) \;\frac{\left( -y\right) ^{n}}{n^{2}}-\frac{1}{2}%
\sum_{n=1}^{\infty }\left( 1-\xi ^{n}\right) \frac{\left( -y\right) ^{n}}{n}%
+O\left( \frac{1}{d}\right) \right)  \label{69}
\end{equation}
that is 
\begin{equation}
Z_{F,th}=\sqrt{\frac{1+y}{1+\xi y}}\exp \left( \frac{d}{\ln \xi ^{-1}}\left[
Li_{+}\left( y\right) -Li_{+}\left( \xi y\right) \right] \right) +O\left( 
\frac{1}{d}\right)  \label{70}
\end{equation}
where 
\begin{equation}
Li_{+}\left( y\right) =\int_{0}^{y}\frac{du}{u}\ln \left( 1+u\right)
=Li_{2}\left( y\right) -\frac{1}{2}Li_{2}\left( y^{2}\right)  \label{71}
\end{equation}
We show in subsection 4.4 that the functions $Li_{2}\left( y\right) $ and $%
Li_{+}\left( y\right) $ can be expressed in terms of the Rogers dilogarithm
function $L(y).$

In the thermodynamic limit, the average number of particles per state is 
\begin{equation}
n_{B}\left( \xi ,y\right) =\frac{\ln \frac{1-\xi y}{1-y}}{\ln \xi ^{-1}}
\label{72}
\end{equation}
\begin{equation}
n_{F}\left( \xi ,y\right) =\frac{\ln \frac{1+y}{1+\xi y}}{\ln \xi ^{-1}}
\label{73}
\end{equation}
Of course, when $\xi \rightarrow 1,$ we recover the average number of
particles per state of section 2. We remind that in formulas (67-70-72-73)$%
,\;y$ is related to $E$ and $\xi y$ is related to $E_{\infty }.$ Finally,
let us remind that $Li_{2}\left( 1\right) =\frac{\pi ^{2}}{6}c_{B}$ and $%
Li_{+}\left( 1\right) =\frac{\pi ^{2}}{6}c_{F}$ where $c_{B}=1$ is the
central charge for bosons, and $c_{F}=1/2$ is the central charge for
fermions.

\bigskip

\subsection{''composite fermions''}

\bigskip

Now, we consider the case of ''composite fermions'' ($l=m>0\;$or$\;g=1).$ We
define 
\begin{equation}
Z_{r}\left( x,y,m,m\right) =\delta _{r,m}+\sum_{p=0}^{E\left( \frac{d-1}{m}%
\right) }x^{\frac{p\left( p-1\right) }{2}m^{2}+prm}\;C_{d+r-1}^{pm+r}\left(
x\right) \;y^{pm+r}  \label{74}
\end{equation}
An easy calculation using (62) shows that 
\begin{equation}
Z_{r}\left( x,y,m,m\right) =\frac{x^{\frac{r\left( m-r\right) }{2}}}{m}%
\sum_{n=0}^{m-1}q^{-rn}\;\left( -q^{n}x^{\frac{1-m}{2}}y;x\right) _{d+r-1}
\label{75}
\end{equation}
where $q=\exp \left( 2i\pi /m\right) .$ The partition function $Z\left(
x,y,m,m\right) $ is obtained from (75) by summing $r$ from $1$ to $m.$

\bigskip

The thermodynamic large $d$-limit when $x=\xi ^{\frac{1}{d}}\rightarrow 1$
can be performed similarly as for the fermion case. Again, the asymptotic
behaviour of (75) is dominated by the term $n=0;$ using (53) and performing
the $\frac{1}{d}$ expansion, we obtain 
\begin{eqnarray}
Z_{r,th}\left( \xi ,y,m,m\right) &=&\frac{\left( 1+\xi y\right) ^{r-1}}{m}%
\left( \frac{1+y}{1+\xi y}\right) ^{\frac{m}{2}}\exp \left( \frac{d}{\ln \xi
^{-1}}\left[ Li_{+}\left( y\right) -Li_{+}\left( \xi y\right) \right] \right)
\notag \\
&&+O\left( \frac{1}{d}\right)  \label{76}
\end{eqnarray}
We may sum the result over $r$ from $1$ to $m$: 
\begin{eqnarray}
Z_{th}\left( \xi ,y,m,m\right) &=&\left( \frac{1+y}{1+\xi y}\right) ^{\frac{m%
}{2}}\frac{\left( 1+\xi y\right) ^{m}-1}{m\xi y}\exp \left( \frac{d}{\ln \xi
^{-1}}\left[ Li_{+}\left( y\right) -Li_{+}\left( \xi y\right) \right] \right)
\notag \\
&&+O\left( \frac{1}{d}\right)  \label{77}
\end{eqnarray}
This result shows that the leading exponential term is the same as for
fermions (70), but here, we obtain some $m$ dependant finite $d$-size
corrections.

\bigskip

\subsection{The partition function Z$\left( x,y,l,1\right) $}

\bigskip

The construction used in this subsection is also valid for $x=1$ and
provides another proof of the result of section 2 without using contour
integrals. Below, we need to specify the number of states $d$.

The partition function is 
\begin{equation}
Z\left( x,y,l,1\right) =Z_{d}\left( y,x\right) =\sum_{N=0}^{E\left( \frac{%
d+l-1}{l}\right) }x^{\frac{N\left( N-1\right) l}{2}}\;C_{d+\left( 1-l\right)
\left( N-1\right) }^{N}\left( x\right) \;y^{N}  \label{78}
\end{equation}
Using the fact that the sum contributes nothing for $N>E\left( \frac{d+l-1}{l%
}\right) $ because of the combinatoric symbol $C,$ and the fact that 
\begin{equation}
C_{n}^{p}\left( x\right) -x^{p}\;C_{n-1}^{p}\left( x\right)
=C_{n-1}^{p-1}\left( x\right)  \label{79}
\end{equation}
we have 
\begin{equation}
Z_{d}\left( y,x\right) -Z_{d-1}\left( xy,x\right) =\sum_{N=1}^{E\left( \frac{%
d+l-1}{l}\right) }x^{\frac{N\left( N-1\right) l}{2}}C_{d+\left( 1-l\right)
\left( N-1\right) -1}^{N-1}\left( x\right) \;\;y^{N}  \label{80}
\end{equation}
Consequently, shifting $N$ by one unity in (80), we just proved the
following recursion relation 
\begin{equation}
Z_{d}\left( y,x\right) =Z_{d-1}\left( xy,x\right) +y\;Z_{d-l}\left(
x^{l}y,x\right)  \label{81}
\end{equation}
This equation (which is also valid for $l=0$ or $1$) makes sense for any
number of states $d$ down to zero provided that we introduce (according to
(78)) 
\begin{eqnarray}
Z_{-1}\left( y,x\right) &=&Z_{-2}\left( y,x\right) =...=Z_{-l+1}\left(
y,x\right) =1  \notag \\
Z_{-l}\left( y,x\right) &=&0  \label{82}
\end{eqnarray}

Now, we suppose $l>1$ and we define 
\begin{equation}
\Psi _{d}\left( y,x\right) =Z_{d}\left( y,x\right)
+\sum_{i=1}^{l-1}W_{i}\left( y,x\right) \;Z_{d-i}\left( x^{i}y,x\right)
\label{83}
\end{equation}
The unknown $d$ independant functions $W_{i}\left( y,x\right) $ are fixed by
the existence of a $d$ independant function $F\left( y,x\right) $ such that 
\begin{equation}
\Psi _{d}\left( y,x\right) =\left[ 1+F\left( y,x\right) \right] \Psi
_{d-1}\left( xy,x\right) =\prod_{i=0}^{d-1}\left[ 1+F\left( x^{i}y,x\right) %
\right] \;\Psi _{0}\left( x^{d}y,x\right)  \label{84}
\end{equation}
The constraints on the functions $W_{i}\left( y,x\right) $ which ensure (83)
and (84) are 
\begin{eqnarray}
W_{1}\left( y,x\right) &=&F\left( y,x\right)  \label{85} \\
\left[ 1+F\left( y,x\right) \right] \;W_{i}\left( xy,x\right)
&=&W_{i+1}\left( y,x\right) \;\;\;\;\;i=1,...,l-2  \label{86} \\
\left[ 1+F\left( y,x\right) \right] \;W_{l-1}\left( xy,x\right) &=&y
\label{87}
\end{eqnarray}
which means that 
\begin{equation}
W_{k}\left( y,x\right) =\prod_{i=0}^{k-2}\left[ 1+F\left( x^{i}y,x\right) %
\right] \;\;F\left( x^{k-1}y,x\right) \;\;\;\;\;\;\;k=2,...,l-1  \label{88}
\end{equation}
\begin{equation}
F\left( x^{l-1}y,x\right) \;\prod_{i=0}^{l-2}\left[ 1+F\left(
x^{i}y,x\right) \right] =y  \label{89}
\end{equation}
Clearly, the last equation is the $x$-deformation of the equivalent equation
(28) of section 2. We know that, for $x$ close enough to $1$, we have $l$
solutions $F_{j}\left( y,x\right) $ which have the property that $%
F_{j}\left( y,1\right) =V_{j}\left( y,l\right) $ described in section 2
below (29).

\bigskip

From (82-83) we have 
\begin{equation}
\Psi _{0}\left( y,x\right) =1+\sum_{i=1}^{l-1}W_{i}\left( y,x\right)
=\prod_{i=0}^{l-2}\left[ 1+F\left( x^{i}y,x\right) \right]  \label{90}
\end{equation}
so that 
\begin{equation}
\Psi _{d}\left( y,x\right) =\prod_{i=0}^{d+l-2}\left[ 1+F\left(
x^{i}y,x\right) \right]  \label{91}
\end{equation}

The partition function is obtained from the linear system (83) ($l$
functions $Z_{d-i}\left( x^{i}y,x\right) $ and $l$ solutions of (89) for $%
\Psi _{d}\left( y,x\right) $ and $W_{i}\left( y,x\right) )$; consequently, 
\begin{equation}
Z(x,y,l,1)=\sum_{j=0}^{l-1}\Delta _{j}\left( y,x\right)
\;\prod_{i=0}^{d+l-2} \left[ 1+F_{j}\left( x^{i}y,x\right) \right]
\label{92}
\end{equation}
\bigskip where $\Delta _{j}\left( y,x\right) $ is the ratio of two
determinants defined by the linear system (83) and where the index $j$ refer
to the $j^{th}$ solution $F_{j}\left( y,x\right) $ of equation (89). We do
not wish to explore further the properties of these determinants (see
appendix B); let us simply remind to the reader that from (29) in section 2,
we know that for $x=1,$ 
\begin{equation}
\Delta _{j}\left( y,1\right) =\frac{1+V_{j}\left( y,l\right) }{%
1+l\;V_{j}\left( y,l\right) }  \label{93}
\end{equation}

\bigskip

We now consider the thermodynamic large $d$-limit of the partition function
with $x=\xi ^{1/d}\rightarrow 1$. As before, for $x$ close enough to $1$ and 
$y$ small enough, we may consider the solution $F_{0}\left( y,x\right) $ to
be dominant in the limit. We must use the following expansion 
\begin{equation}
\ln \left[ 1+F\left( y,x\right) \right] =\sum_{n=1}^{\infty }a_{n}\left(
x\right) \;y^{n}  \label{94}
\end{equation}
together with the property 
\begin{equation}
\int_{0}^{y}\frac{dv}{v}\ln \left[ 1+F\left( v,x\right) \right]
=\sum_{n=1}^{\infty }a_{n}\left( x\right) \;\frac{y^{n}}{n}  \label{95}
\end{equation}
We may now make a $\frac{1}{d}$ expansion of $\prod_{i=0}^{d+l-2}\left[
1+F\left( \xi ^{\frac{i}{d}}y,\xi ^{\frac{1}{d}}\right) \right] $ and write 
\begin{eqnarray}
Z_{th}\left( \xi ,y,l,1\right) &=&\left( \frac{1+V\left( y,l\right) }{%
1+V\left( \xi y,l\right) }\right) ^{\frac{3}{2}}\frac{\left[ 1+V\left( \xi
y,l\right) \right] ^{l}}{1+l\;V\left( y,l\right) }\exp \left[ \Phi \left(
\xi y\right) -\Phi \left( y\right) \right] .  \notag \\
&&.\exp \left( \frac{d}{\ln \xi ^{-1}}\int_{\xi y}^{y}\frac{dv}{v}\ln \left[
1+V\left( v,l\right) \right] \right) +O\left( \frac{1}{d}\right)  \label{96}
\end{eqnarray}
where 
\begin{equation}
\Phi \left( y\right) =\left[ \frac{\partial }{\partial x}\int_{0}^{y}\frac{dv%
}{v}\ln \left[ 1+F_{0}\left( v,x\right) \right] \right] _{x=1}  \label{97}
\end{equation}
This ends the calculation of the finite $d$-size corrections in the case $%
m=1.$ At the end of subsection 4.4, we show that the integral in (96) can be
expressed in terms of the functions $Li_{+},\;Li_{2}$ or of the Rogers
dilogarithm function.

\bigskip

\subsection{\protect\bigskip The partition function Z$\left( x,y,l,m\right) $%
}

The partition function in the general case (for any $l$ and $m$) is
calculated in appendix B.\ We simply give the result and its thermodynamic
limit. We obtain 
\begin{eqnarray}
Z_{r}\left( x,y,l,m\right) &=&\frac{x^{rd+\frac{r\left( r-1\right) }{2}}y^{r}%
}{m}\sum_{j=0}^{\sup \left( m,l\right) -1}\Delta _{j}\left( y,x\right)
\prod_{i=0}^{d+r-2}\left[ 1+F_{j}\left( x^{i}y,x\right) \right] .  \notag \\
&&.\prod_{i=0}^{r-1}\left[ F_{j}\left( x^{d+i-1+\left( r-i\right) \frac{l}{m}%
}y,x\right) \right] ^{-1}  \label{98}
\end{eqnarray}
In this expression, $\Delta _{j}\left( y,x\right) $ is the ratio of two
determinants and is explained in Appendix B; the functions $F_{j}\left(
y,x\right) $ satisfy the equations 
\begin{equation}
x^{-\left( m-l\right) }\prod_{i=0}^{m-1}F\left( x^{\frac{i}{m}}y,x\right)
=x^{-\frac{\left( m-1\right) \left( l-1\right) }{2}}\;y^{m}\;%
\prod_{i=0}^{m-l-1}\left[ 1+F\left( x^{\frac{i}{m}}y,x\right) \right]
\;\;\;m\geq l  \label{99}
\end{equation}
\begin{equation}
\prod_{i=0}^{m-1}F\left( x^{\frac{l-m+i}{m}}y,x\right) \;\prod_{i=0}^{l-m-1}%
\left[ 1+F\left( x^{\frac{i}{m}}y,x\right) \right] =x^{-\frac{\left(
m-1\right) \left( l-1\right) }{2}}\;y^{m}\;\;\;l\geq m  \label{100}
\end{equation}
where $\prod_{0}^{-1}\left[ ...\right] =1$ by convention. The equations
(98), (99) and(100) are the $x$-deformed equations (A18), (A15a) and (A15b)
of appendix A. By identification of (98) when $x=1$ with (A18), we see that 
\begin{equation}
F_{j}\left( y,1\right) =V_{j}  \label{101}
\end{equation}
\begin{equation}
\Delta _{j}\left( y,1\right) =\frac{m\left( 1+V_{j}\right) }{m+l\;V_{j}}
\label{102}
\end{equation}

We are now taking the thermodynamic limit when $x=\xi ^{\frac{1}{d}%
}\rightarrow 1$ and when the number of states $d\rightarrow \infty .$ The
technique is exactly similar to the one used in subsection 4.3; we obtain 
\begin{eqnarray}
Z_{r,th}\left( \xi ,y,l,m\right) &=&\left[ 1+V\left( \xi y,l/m\right) \right]
^{\frac{rl}{m}}\left( \frac{1+V\left( y,l/m\right) }{1+V\left( \xi
y,l/m\right) }\right) ^{3/2}\frac{\exp \left[ \Phi \left( \xi y\right) -\Phi
\left( y\right) \right] }{m+l\;V\left( y,l/m\right) }.  \notag \\
&&.\exp \left( \frac{d}{\ln \xi ^{-1}}\int_{\xi y}^{y}\frac{dv}{v}\ln \left[
1+V\left( v,l/m\right) \right] \right) +O\left( \frac{1}{d}\right)
\label{103}
\end{eqnarray}
where 
\begin{equation}
\Phi \left( y\right) =\left[ \frac{\partial }{\partial x}\int_{0}^{y}\frac{dv%
}{v}\ln \left[ 1+F_{0}\left( v,x\right) \right] \right] _{x=1}  \label{104}
\end{equation}
Finally, the partition function in the thermodynamic limit is obtained from
(103) by summing $r$ from $1$ to $m$%
\begin{eqnarray}
Z_{th}\left( \xi ,y,l,m\right) &=&\frac{\left[ 1+V\left( \xi y,l/m\right) %
\right] ^{l}-1}{1-\left[ 1+V\left( \xi y,l/m\right) \right] ^{-l/m}}\left( 
\frac{1+V\left( y,l/m\right) }{1+V\left( \xi y,l/m\right) }\right) ^{3/2}%
\frac{\exp \left[ \Phi \left( \xi y\right) -\Phi \left( y\right) \right] }{%
m+l\;V\left( y,l/m\right) }.  \notag \\
&&.\exp \left( \frac{d}{\ln \xi ^{-1}}\int_{\xi y}^{y}\frac{dv}{v}\ln \left[
1+V\left( v,l/m\right) \right] \right) +O\left( \frac{1}{d}\right)
\label{105}
\end{eqnarray}
Using (37), the integral in (105) can be transformed into $Li_{+}$ and $%
Li_{2}\;$functions since 
\begin{eqnarray}
\int_{0}^{y}\frac{dv}{v}\ln \left[ 1+V\left( v,l/m\right) \right] &=&Li_{+}%
\left[ V\left( y,l/m\right) \right] +\frac{l-m}{2m}\ln ^{2}\left[ 1+V\left(
y,l/m\right) \right]  \label{106} \\
&=&Li_{2}\left[ \frac{V\left( y,l/m\right) }{1+V\left( y,l/m\right) }\right]
+\frac{l}{2m}\ln ^{2}\left[ 1+V\left( y,l/m\right) \right]  \notag \\
&&  \label{107}
\end{eqnarray}
and finally using the equation 
\begin{equation}
V\left( 1+V\right) ^{\frac{l-m}{m}}=y  \label{108}
\end{equation}
we also have 
\begin{equation}
\int_{0}^{y}\frac{dv}{v}\ln \left[ 1+V\left( v,l/m\right) \right] =L\left[ 
\frac{V\left( y,l/m\right) }{1+V\left( y,l/m\right) }\right] +\frac{1}{2}\ln
y\;.\;\ln \left[ 1+V\left( y,l/m\right) \right]  \label{109}
\end{equation}
where the function $L\left( y\right) $ is the Rogers dilogarithm function 
\begin{equation}
L\left( y\right) =-\frac{1}{2}\int_{0}^{y}dv\left[ \frac{\ln v}{1-v}+\frac{%
\ln \left( 1-v\right) }{v}\right]  \label{110}
\end{equation}
This result has been obtained by Berkovich and McCoy$^{\left[ 12\right] }.$
The difference between their result and ours in (105) is:

1$%
{{}^\circ}%
)$ their result is obtained using the steepest descent method and provides
only (109) while we get the finite $d$-size corrections,

2$%
{{}^\circ}%
)$ they calculated the limit $x\rightarrow 1\;\prod_{i=0}^{\infty }\left[
1+F\left( x^{i}y,x\right) \right] \;$while we calculated the limit $%
d\rightarrow \infty \;\prod_{i=0}^{d-1}\left[ 1+F\left( \xi ^{\frac{i}{d}%
}y,\xi ^{\frac{1}{d}}\right) \right] $ which generates the $\xi $ dependant
subtracted terms.

\bigskip

In our limit, the average number of particles per state is found to be 
\begin{equation}
n\left( \xi ,y\right) =\frac{\ln \left( \frac{1+V\left( y,g\right) }{%
1+V\left( \xi y,g\right) }\right) }{\ln \xi ^{-1}}  \label{111}
\end{equation}
where $g=l/m.$

From conformal field theory, we know that the central charge is given by 
\begin{equation}
c=\frac{6}{\pi ^{2}}\int_{0}^{1}\frac{dv}{v}\ln \left[ 1+V\left(
v,l/m\right) \right] =\frac{L\left[ \frac{V\left( 1,g\right) }{1+V\left(
1,g\right) }\right] }{L(1)}  \label{112}
\end{equation}
$\;$Following Hikami$^{\left[ 11\right] }$ we define 
\begin{equation}
\xi _{H}^{g}=\frac{V\left( 1,g\right) }{1+V\left( 1,g\right) }  \label{113}
\end{equation}
and we obtain the central charge as 
\begin{equation}
c=\frac{L\left( \xi _{H}^{g}\right) }{L\left( 1\right) }  \label{114}
\end{equation}
where $\xi _{H}$ satisfy 
\begin{equation}
\xi _{H}^{g}=1-\xi _{H}  \label{115}
\end{equation}

\bigskip

\bigskip

\textbf{Acknowledgments:\ \ }I wish to thank M.\ Bauer who performed several
calculations on MACSYMA in order to check the validity and the general
structure of the results on the partition function.\ He also mentionned to
me the contour integral representation of the combinatoric symbol that he
used himself in Ref.$\left[ 14\right] .\;$I thank him again for a careful
reading of the manuscript.

\bigskip

\bigskip

\bigskip

\section{\protect\bigskip Appendix A}

\bigskip

\bigskip 1$%
{{}^\circ}%
)\;$\textbf{The function }$V\left( y,\alpha \right) $

\bigskip We define the function $V\left( y,\alpha \right) $ as the solution
of the equation 
\begin{equation}
z\left( 1+z\right) ^{\alpha -1}=y\;\;\;\;\alpha \in \;R  \tag{A1}
\end{equation}
which has a small $y$ expansion of the type 
\begin{equation}
V\left( y,\alpha \right) =y+\left( 1-\alpha \right) \;y^{2}+O\left(
y^{3}\right)  \tag{A2}
\end{equation}
We may calculate the complete $y$ expansion of $V^{k}\left( y,\alpha \right) 
$ for any integer $k$ from the contour integral 
\begin{equation}
V^{k}\left( y,\alpha \right) =\frac{1}{2i\pi }\oint_{C_{V}}dz\;\frac{%
z^{k}\left( 1+z\right) ^{\alpha -2}\left( 1+\alpha z\right) }{z\left(
1+z\right) ^{\alpha -1}-y}  \tag{A3}
\end{equation}
where the numerator is $z^{k}$ times the derivative of the denominator and
where the contour $C_{V}$ surrounds the point $z=V\left( y,\alpha \right) .$
In order to obtain the formal $y$ expansion, we deform the contour to
include the point $z=0$ but exclude the point $z=-1.$ We easily obtain 
\begin{eqnarray}
V^{k}\left( y,\alpha \right) &=&y^{k}+k\left( 1-\alpha \right) y^{k+1}+ 
\notag \\
&&+k\left( \alpha -1\right) \sum_{n=2}^{\infty }\left( -\right) ^{n}\frac{%
\Gamma \left[ n\alpha +k\left( \alpha -1\right) \right] }{\Gamma \left[
n\alpha +k\left( \alpha -1\right) -n+1\right] }\frac{y^{k+n}}{n!} 
\label{A4}
\end{eqnarray}
This expansion is in fact valid for any $k\in R$ and for instance for $k=-1,$
we obtain the $y$ expansion (5) for the function $W(y,g)=V^{-1}\left(
y,g\right) $ which enters the expression (3) for the average number of
particles \ per site. From the above expansion and from the \ equation (A1),
it is easy to obtain some other useful expansions like 
\begin{equation}
\left( 1+V\left( y,\alpha \right) \right) ^{k}=1+ky-k\sum_{n=2}^{\infty
}\left( -\right) ^{n}\frac{\left[ n\alpha -k-1\right] ...\left[ n\alpha
-k-n+1\right] }{n!}\;y^{n}\;\;\;  \tag{A5}
\end{equation}
and 
\begin{equation}
\ln \left( 1+V\left( y,\alpha \right) \right) =y-\sum_{n=2}^{\infty }\left(
-\right) ^{n}\frac{\left( n\alpha -1\right) ...\left( n\alpha -n+1\right) }{%
n!}\;\;y^{n}  \tag{A6}
\end{equation}

\bigskip

\bigskip

We now consider the solutions of the polynomial equation 
\begin{equation}
z^{m}=\left( 1+z\right) ^{m-l}\;y^{m}\;\;\;\;\;\;\;\;\;m\geq l  \tag{A7}
\end{equation}
where $m$ and $l$ are non negative integers. The above polynomial in $z\;$%
has $m$ roots which are 
\begin{equation}
z_{j}=V\left( q^{j}y,l/m\right) \;\;\;\ j=0,...,m-1  \tag{A8}
\end{equation}
where $q=\exp \left( 2i\pi /m\right) .$

\bigskip

We now consider the solutions of the polynomial equation 
\begin{equation}
z^{m}\left( 1+z\right) ^{l-m}=y^{m}\;\;\;\;\;\;\;\;\;l>m  \tag{A9}
\end{equation}
where $m$ and $l$ are non negative integers. The above polynomial in $z\;$%
has $l$ roots, $m$ of which $\rightarrow 0$ when $y\rightarrow 0;$ they are 
\begin{equation}
z_{j}=V\left( q^{j}y,l/m\right) \;\;\;\;\;\;j=0,...,m-1  \tag{A10}
\end{equation}
where $q=\exp \left( 2i\pi /m\right) .$ The $\left( l-m\right) $ remaining
roots $\rightarrow -1$ when $y\rightarrow 0.$ If we write $z=-1-t,$ we
obtain for $t$ a similar equation as (A9). Consequently, the remaining roots
can be written as 
\begin{equation}
z_{j}=-1-V\left( Q^{2j+l}\;y^{\frac{m}{l-m}},\frac{l}{l-m}\right)
\;\;\;\;\;j=m,...,l-1  \tag{A11}
\end{equation}
where $Q=\exp \left( i\pi /\left( l-m\right) \right) .$

\bigskip

\bigskip

2$%
{{}^\circ}%
)\;$\textbf{Calculation of the partition function }$Z(y,l,m).$

\bigskip

We define 
\begin{equation}
Z_{r}\left( y,l,m\right) =\delta _{r,m}+\sum_{p=0}^{E\left( \frac{d-1}{l}%
\right) }\;C_{d+p\left( m-l\right) +r-1}^{pm+r}\;\;\;y^{pm+r}  \tag{A12}
\end{equation}
Using the contour integral representation (26) for the combinatoric symbol,
we write 
\begin{equation}
Z_{r}\left( y,l,m\right) =\delta _{r,m}+\sum_{p=0}^{E\left( \frac{d-1}{l}%
\right) }\frac{1}{2i\pi }\oint_{C_{0}}dz\;\frac{\left( 1+z\right)
^{d+p\left( m-l\right) +r-1}}{z^{pm+r+1}}\;\;y^{pm+r}  \tag{A13}
\end{equation}
where $C_{0}$ surrounds the point $z=0.$ The sum over $p$ can be extended to 
$\infty $ since the contour integral gives $0$ for $p>E\left( \frac{d-1}{l}%
\right) .$ We obtain 
\begin{equation}
Z_{r}\left( y,l,m\right) =\frac{y^{r}}{2i\pi }\oint_{C_{l,m}}dz\;\frac{%
z^{m-r-1}\left( 1+z\right) ^{d+r-1}}{z^{m}-\left( 1+z\right) ^{m-l}\;y^{m}} 
\tag{A14}
\end{equation}
The convergence of the series requires $\left| z\right| ^{m}\left|
1+z\right| ^{l-m}>y$ \ which is certainly true for large $\left| z\right| .$
Consequently, the contour $C_{l,m}$ surrounds all the roots of the
polynomial equation 
\begin{equation}
z^{m}-\left( 1+z\right) ^{m-l}\;y^{m}=0\;\;\;\;\;m\geq l  \tag{A15a}
\end{equation}
or 
\begin{equation}
z^{m}\left( 1+z\right) ^{l-m}-y^{m}=0\;\;\;\;\;l\geq m  \tag{A15b}
\end{equation}
In (A14), the extra pole at $z=0$ when $r=m$ has been taken into account to
cancel the term $\delta _{r,m}$ in (A13);\ in that case, the contour $%
C_{l,m} $ does not surround the point $z=0.$ The equation (A15a) has $m$
roots 
\begin{equation}
V_{j}=V\left( q^{j}y,l/m\right) \;\;\;\;\;\;\;j=0,...,m-1  \tag{A16}
\end{equation}
The equation (A15b) has $l$ roots 
\begin{eqnarray}
V_{j} &=&V\left( q^{j}y,l/m\right) \;\;\;\;\;\;j=0,...,m-1  \notag \\
V_{j} &=&-1-V\left( Q^{2j+l}\;y^{\frac{m}{l-m}},\frac{l}{l-m}\right)
\;\;\;\;\;j=m,...,l-1  \label{A17}
\end{eqnarray}
with $q=\exp \left( 2i\pi /m\right) $ and $Q=\exp \left( i\pi /\left(
l-m\right) \right) .$ \ The contour integral (A14) is now calculated by
summing over all roots 
\begin{equation}
Z_{r}\left( y,l,m\right) =\sum_{j=0}^{Sup\left( m,l\right) -1}\;\frac{\left(
1+V_{j}\right) ^{d+r}}{m+l\;V_{j}}\;\frac{y^{r}}{V_{j}^{r}}\;  \tag{A18}
\end{equation}
Finally the partition function is obtained by summing the geometrical series
in (A18) from $r=1$ to $r=m:$%
\begin{equation}
Z\left( y,l,m\right) =\sum_{j=0}^{Sup\left( m,l\right) -1}\;\frac{\left(
1+V_{j}\right) ^{l}-1}{\left( 1+V_{j}\right) ^{\frac{l}{m}}-q^{k\left(
j\right) }}\;\frac{\left( 1+V_{j}\right) ^{d+\frac{l}{m}}}{m+l\;V_{j}} 
\tag{A19}
\end{equation}
where $k\left( j\right) \;$is an integer equal to$\;j$ for $j=0,...,m-1$ but
depends upon the definition of $\left( 1+V_{j}\right) ^{1/m}$ in (A17) for $%
l>m$ and $j=m,...,l-1.$

\section{Appendix B}

\bigskip

In this appendix, we calculate the partition function 
\begin{equation}
Z_{r}\left( x,y,l,m\right) =Z_{d,r}\left( y,x\right) =\delta
_{r,m}+\sum_{p=0}^{E\left( \frac{d-1}{l}\right) }x^{\left( \frac{p\left(
p-1\right) }{2}m+pr\right) l}\;C_{d+p\left( m-l\right) +r-1}^{pm+r}\left(
x\right) \;y^{pm+r}  \tag{B1}
\end{equation}
for any value of $l$ and $m.$ The method used here is the same as in section
4.3 for $m=1;$ we calculate the difference $Z_{d,r}\left( y,x\right)
-Z_{d-1,r}\left( xy,x\right) $ and we use the relation (79) between the
combinatoric symbols $C.$ The result depends on the value of $r:$%
\begin{equation}
Z_{d,r}\left( y,x\right) =Z_{d-1,r}\left( xy,x\right) +x^{-\left( r-1\right) 
\frac{l}{m}}\;y\;Z_{d,r-1}\left( x^{\frac{l}{m}}y\right)
\;\;\;\;\;\;r=2,3,...,m  \tag{B2}
\end{equation}
\begin{equation}
Z_{d,1}\left( y,x\right) =Z_{d-1,1}\left( xy,x\right) +y\;Z_{d-l,m}\left( x^{%
\frac{l}{m}}y\right)  \tag{B3}
\end{equation}
In order to use these recursions we may introduce a generalized Fourier
transform which diagonalizes the system over $r$ and in the same time
performs subtractions over the number of states $d$. As a matter of fact,
the subtractions over $d$ are rational so that we must consider the
existence of $Z_{d,r}\left( y,x\right) $ for any rational $d$. We define 
\begin{equation}
T_{d,n}\left( y,x\right) =\sum_{r=1}^{m}q^{nr}\;x^{-\frac{r\left( m-r\right)
l}{2m}}\;Z_{d-\left( r-1\right) \frac{l}{m},r}\left( y,x\right)
\;\;\;\;n=0,...,m-1  \tag{B4}
\end{equation}
where $q=\exp (2i\pi /m).$ The generalized Fourier transform (B4) can be
inverted as 
\begin{equation}
Z_{d,r}\left( y,x\right) =\frac{x^{\frac{r\left( m-r\right) l}{2m}}}{m}%
\sum_{n=0}^{m-1}q^{-nr}\;T_{d+\left( r-1\right) \frac{l}{m},n}\left(
y,x\right) \;\;\;r=1,...,m  \tag{B5}
\end{equation}
Then, the recursions (B2) and (B3) are equivalent to the following recursion 
\begin{equation}
T_{d,n}\left( y,x\right) =T_{d-1,n}\left( xy,x\right) +q^{n}\;x^{-\frac{%
\left( m-1\right) l}{2m}}\;y\;T_{d-\frac{l}{m},n}\left( x^{\frac{l}{m}%
}y,x\right)  \tag{B6}
\end{equation}
The solution of the above recursion is possible if we introduce two coprime
integers $l^{\prime }$ and $m^{\prime }$ such that $\frac{l}{m}=\frac{%
l^{\prime }}{m^{\prime }}.$ \ It is important all along the solution to keep
track of the $q^{n}$ terms so that we may at the end perform the inverse
Fourier transform (B5). Let us decompose $n$ in a unique way (Bezout
decomposition) as 
\begin{equation}
n=n^{\prime }+sl\;(\func{mod}m)\;\;\;\;0\leq n^{\prime }\leq \frac{m}{%
m^{\prime }}-1\;\;\;\;\;0\leq s\leq m^{\prime }-1  \tag{B7}
\end{equation}
Now, the solution of the above recursion depends upon the relative values of 
$l$ and $m.$

1$%
{{}^\circ}%
)\;m>l$

\bigskip

We define 
\begin{equation}
\Psi _{d,n}\left( y,x\right) =T_{d,n}\left( y,x\right)
+\sum_{i=1}^{m^{\prime }-1}Q^{is}\;W_{i,n^{\prime }}\left( y,x\right) \;T_{d-%
\frac{i}{m^{\prime }},n}\left( x^{\frac{i}{m^{\prime }}}y,x\right)  \tag{B8}
\end{equation}
where $Q=\exp \left( 2i\pi /m^{\prime }\right) \;$and where the functions $%
W_{i,n^{\prime }}\left( y,x\right) $ are determined by the condition 
\begin{equation}
\Psi _{d,n}\left( y,x\right) =Q^{s}\;U_{n^{\prime }}\left( y,x\right) \;\Psi
_{d-\frac{1}{m^{\prime }},n}\left( x^{\frac{1}{m^{\prime }}}y,x\right) 
\tag{B9}
\end{equation}
which implies 
\begin{equation}
\Psi _{d,n}\left( y,x\right) =\prod_{i=0}^{m^{\prime }-1}U_{n^{\prime
}}\left( x^{\frac{i}{m^{\prime }}}y,x\right) \;\Psi _{d-1,n}\left(
xy,x\right)  \tag{B10}
\end{equation}
The constraints (B8) and (B9) gives the relations 
\begin{eqnarray}
W_{k,n^{\prime }}\left( y,x\right) &=&U_{n^{\prime }}\left( y,x\right)
\;W_{k-1,n^{\prime }}\left( x^{\frac{1}{m^{\prime }}}y,x\right)
\;\;\;\;\;\;\;k=2,...,l^{\prime }-1,l^{\prime }+1,...,m^{\prime }-1  \notag
\\
W_{1,n^{\prime }}\left( y,x\right) &=&U_{n^{\prime }}\left( y,x\right)
\;\;\;\;\;if\;l^{\prime }\neq 1  \label{B11a}
\end{eqnarray}
\begin{equation}
q^{n^{\prime }}\;x^{-\frac{\left( m-1\right) l}{2m}}\;y+W_{l^{\prime
},n^{\prime }}\left( y,x\right) =U_{n^{\prime }}\left( y,x\right)
\;W_{l^{\prime }-1,n^{\prime }}\left( x^{\frac{1}{m^{\prime }}}y,x\right) 
\tag{B11b}
\end{equation}
where eventually $W_{0,n^{\prime }}\left( y,x\right) =1,$%
\begin{equation}
1=U_{n^{\prime }}\left( y,x\right) \;W_{m^{\prime }-1,n^{\prime }}\left( x^{%
\frac{1}{m^{\prime }}}y,x\right)  \tag{B11c}
\end{equation}
These equations can be solved recursively; we obtain 
\begin{equation}
W_{k,n^{\prime }}\left( y,x\right) =\prod_{i=0}^{k-1}U_{n^{\prime }}\left(
x^{\frac{i}{m^{\prime }}}y,x\right) \;\;\;\;k=1,...,l^{\prime }-1  \tag{B12a}
\end{equation}
\begin{equation}
W_{k,n^{\prime }}\left( y,x\right) =\prod_{i=0}^{k-l^{\prime
}-1}U_{n^{\prime }}\left( x^{\frac{i}{m^{\prime }}}y,x\right) \;W_{l^{\prime
},n^{\prime }}\left( x^{\frac{k-l^{\prime }}{m^{\prime }}}y,x\right)
\;\;\;\;\;k=l^{\prime }+1,...,m^{\prime }-1  \tag{B12b}
\end{equation}
\begin{equation}
q^{n^{\prime }}\;x^{-\frac{\left( m-1\right) l}{2m}}\;y+W_{l^{\prime
},n^{\prime }}\left( y,x\right) =\prod_{i=0}^{l^{\prime }-1}U_{n^{\prime
}}\left( x^{\frac{i}{m^{\prime }}}y,x\right)  \tag{B12c}
\end{equation}
\begin{equation}
1=\prod_{i=0}^{m^{\prime }-l^{\prime }-1}U_{n^{\prime }}\left( x^{\frac{i}{%
m^{\prime }}}y,x\right) \;W_{l^{\prime },n^{\prime }}\left( x^{\frac{%
m^{\prime }-l^{\prime }}{m^{\prime }}}y,x\right)  \tag{B12d}
\end{equation}
If we transform (B12c) by changing $\ y\rightarrow x^{\frac{m^{\prime
}-l^{\prime }}{m^{\prime }}}\;y\;$and if we multiply by $\prod_{i=0}^{m^{%
\prime }-l^{\prime }-1}U_{n^{\prime }}\left( x^{\frac{i}{m^{\prime }}%
}y,x\right) $ we obtain from (B12d) 
\begin{equation}
q^{n^{\prime }}\;x^{-\frac{\left( m-1\right) l}{2m}}\;y\;\prod_{i=0}^{m^{%
\prime }-l^{\prime }-1}U_{n^{\prime }}\left( x^{\frac{i}{m^{\prime }}%
}y,x\right) =x^{-\frac{m^{\prime }-l^{\prime }}{m^{\prime }}}\;F_{n^{\prime
}}\left( y,x\right)  \tag{B13}
\end{equation}
where $1+F_{n^{\prime }}(y,x)=\prod_{i=0}^{m^{\prime }-1}U_{n^{\prime
}}\left( x^{\frac{i}{m^{\prime }}}y,x\right) .$\ Again, if we multiply the
above equation by $\left( m^{\prime }-1\right) $ similar equations, each
being deduced from the precedent one by the transformation $y\rightarrow x^{%
\frac{1}{m^{\prime }}}y,$ we obtain 
\begin{eqnarray}
x^{-\left( m^{\prime }-l^{\prime }\right) }\prod_{i=0}^{m^{\prime
}-1}F_{n^{\prime }}\left( x^{\frac{i}{m^{\prime }}}y,x\right)
&=&q^{n^{\prime }m^{\prime }}\;x^{-\frac{\left[ \left( m-1\right) l^{\prime
}-m^{\prime }+1\right] }{2}}.  \notag \\
&&.\;y^{m^{\prime }}\;\prod_{i=0}^{m^{\prime }-l^{\prime }-1}\left[
1+F_{n^{\prime }}\left( x^{\frac{i}{m^{\prime }}}y,x\right) \right] 
\label{B14a}
\end{eqnarray}
If we multiply (B14a) by $m/m^{\prime }$ different equations, each being
deduced from the precedent one by the transformation $y\rightarrow x^{\frac{1%
}{m}}y,$ we obtain 
\begin{equation}
x^{-\left( m-l\right) }\prod_{i=0}^{m-1}F_{n^{\prime }}\left( x^{\frac{i}{m}%
}y,x\right) =x^{-\frac{\left( m-1\right) \left( l-1\right) }{2}%
}\;y^{m}\;\prod_{i=0}^{m-l-1}\left[ 1+F_{n^{\prime }}\left( x^{\frac{i}{m}%
}y,x\right) \right]  \tag{B14b}
\end{equation}
which is the $x$-deformed equation (A15a). At $x=1,$ equation (B14a) is
polynomial of degree $m^{\prime }.$ For a given $n^{\prime },$ we have $%
m^{\prime }$ solutions $F_{n^{\prime }}^{\left( j\right) },$ $%
j=1,...,m^{\prime }$; this analysis remain true by $x$ deformation for $x$
not too far from $1.$ On the other hand, equation (B14b) is $n^{\prime }$
independant and has $m$ solutions $F_{u}\left( y,x\right) $ which are the
same globally as the solutions $F_{n^{\prime }}^{\left( j\right) }$ for all $%
n^{\prime }$ and all $j.\;$Considering for a given $n^{\prime },\;$the $%
m^{\prime }$ solutions $F_{n^{\prime }}^{\left( j\right) },$ $%
j=1,...,m^{\prime },$ the system (B8) is a linear system of $m^{\prime }$
equations for $m^{\prime }$ unknown functions $T_{d-\frac{i}{m^{\prime }}%
}\left( x^{\frac{i}{m^{\prime }}}y,x\right) $\ $,\;i=0,...,m^{\prime }-1.$
Consequently, we may write 
\begin{equation}
T_{d-\frac{i}{m^{\prime }}}\left( x^{\frac{i}{m^{\prime }}}y,x\right)
=Q^{-is}\;\sum_{j=1}^{m^{\prime }}\Delta _{i,\left( j\right) }^{n^{\prime
}}\left( y,x\right) \;\Psi _{d,n}^{\left( j\right) }\left( y,x\right)
\;\;\;i=0,...,m^{\prime }-1  \tag{B15}
\end{equation}
where $\Delta _{i,\left( j\right) }^{n^{\prime }}\left( y,x\right) $ is the
ratio of two determinants containing only the functions $W_{i,n^{\prime
}}^{\left( j\right) }\left( y,x\right) ,$ the powers of $Q$ being extracted
separately. The functions $\Delta _{i,\left( j\right) }^{n^{\prime }}\left(
y,x\right) $ satisfy the relations 
\begin{equation}
\Delta _{i,\left( j\right) }^{n^{\prime }}\left( y,x\right) =W_{m^{\prime
}-1,n^{\prime }}^{\left( j\right) }\left( x^{\frac{1}{m^{\prime }}%
}y,x\right) \;\Delta _{i-1,\left( j\right) }^{n^{\prime }}\left( x^{\frac{1}{%
m^{\prime }}}y,x\right) \;\;\;\;\;i=1,...,m^{\prime }-1  \tag{B16}
\end{equation}
and the orthogonality relations 
\begin{eqnarray}
\sum_{i=0}^{m^{\prime }-1}W_{i,n^{\prime }}^{\left( k\right) }\left(
y,x\right) \;\Delta _{i,\left( j\right) }^{n^{\prime }}\left( y,x\right)
&=&\delta _{j}^{k}  \notag \\
\sum_{j=1}^{m^{\prime }}W_{i,n^{\prime }}^{\left( j\right) }\left(
y,x\right) \;\Delta _{k,\left( j\right) }^{n^{\prime }}\left( y,x\right)
&=&\delta _{i,k}  \label{B17}
\end{eqnarray}
Finally, using 
\begin{equation}
\Psi _{d,n}^{\left( j\right) }\left( y,x\right) =\prod_{i=0}^{d-1}\left[
1+F_{n^{\prime }}^{\left( j\right) }\left( x^{i}y,x\right) \right] \;\Psi
_{0,n}^{\left( j\right) }\left( x^{d}y,x\right)  \tag{B18}
\end{equation}
the solution of the recursion (B6) is 
\begin{equation}
T_{d-\frac{i}{m^{\prime }}}\left( x^{\frac{i}{m^{\prime }}}y,x\right)
=Q^{-is}\;\sum_{j=1}^{m^{\prime }}\Delta _{i,\left( j\right) }^{n^{\prime
}}\left( y,x\right) \;\prod_{k=0}^{d-1}\left[ 1+F_{n^{\prime }}^{\left(
j\right) }\left( x^{k}y,x\right) \right] \;\Psi _{0,n}^{\left( j\right)
}\left( x^{d}y,x\right)  \tag{B19}
\end{equation}
We are now in position to calculate the inverse Fourier transform (B5). We
write $(r-1)l^{\prime }=Km^{\prime }-h$ with $K$ integer and $0\leq
h<m^{\prime };$ then, we find 
\begin{eqnarray}
Z_{d,r}\left( y,x\right) &=&\frac{x^{\frac{r\left( m-r\right) l}{2m}}}{m}%
\sum_{n^{\prime },j}q^{-n^{\prime }r}\Delta _{h,\left( j\right) }^{n^{\prime
}}\left( x^{-\frac{h}{m^{\prime }}}y,x\right) .  \notag \\
&&.\prod_{k=0}^{d+K-1}\left[ 1+F_{n^{\prime }}^{\left( j\right) }\left( x^{k-%
\frac{h}{m^{\prime }}}y,x\right) \right] \Phi _{n^{\prime }}^{\left(
j\right) }\left( x^{d+\frac{\left( r-1\right) l^{\prime }}{m^{\prime }}%
}y,x\right)  \label{B20}
\end{eqnarray}
where the sums over $n^{\prime }$ and $j$ runs respectively from $0$ to $%
\frac{m}{m^{\prime }}-1$ and from $1$ to $m^{\prime };$ in (B20), we
introduced the functions 
\begin{equation}
\Phi _{n^{\prime }}^{\left( j\right) }\left( y,x\right)
=\sum_{s=0}^{m^{\prime }-1}Q^{-sl^{\prime }}\Psi _{0,n}^{\left( j\right)
}\left( y,x\right)  \tag{B21}
\end{equation}
which may be determined from the values of $Z_{d,r}\left( y,x\right) $ in
(B1) for $0\leq d<l+1.$ We find 
\begin{equation}
\Phi _{n^{\prime }}^{\left( j\right) }\left( y,x\right) =W_{l^{\prime
},n^{\prime }}^{\left( j\right) }\left( y,x\right)  \tag{B22}
\end{equation}
Now, the expression \ (B20) can be transformed using successively (B16),
(B11c), (B22), (B12d) and the expression of $\left[ 1+F_{n^{\prime }}\right] 
$ in terms of $U_{n^{\prime }}$; we obtain 
\begin{eqnarray}
Z_{d,r}\left( y,x\right) &=&\frac{x\frac{r\left( m-r\right) l}{2m}}{m}%
\sum_{n^{\prime },j}q^{-n^{\prime }r}\Delta _{0,\left( j\right) }^{n^{\prime
}}\left( y,x\right) \prod_{k=0}^{d+r-2}\left[ 1+F_{n^{\prime }}^{\left(
j\right) }\left( x^{k}y,x\right) \right] .  \label{B23} \\
&&.\prod_{i=0}^{r\left( m^{\prime }-l^{\prime }\right) -1}W_{m^{\prime
}-1,n^{\prime }}^{\left( j\right) }\left( x^{d-1+\frac{rl^{\prime }+i+1}{%
m^{\prime }}}y,x\right)  \notag
\end{eqnarray}
Finally, using (B13) $r$ times and (B11c), we get 
\begin{eqnarray}
Z_{d,r}\left( y,x\right) &=&\frac{x^{rd+\frac{r\left( r-1\right) }{2}}y^{r}}{%
m}\sum_{u=0}^{m-1}\Delta _{u}\left( y,x\right) \prod_{k=0}^{d+r-2}\left[
1+F_{u}\left( x^{k}y,x\right) \right] .  \notag \\
&&.\prod_{i=0}^{r-1}\left[ F_{u}\left( x^{d+i-1+\left( r-i\right) \frac{l}{m}%
}y,x\right) \right] ^{-1}  \label{B24}
\end{eqnarray}
where $u=n^{\prime }+\left( j-1\right) \frac{m}{m^{\prime }}.$ Equation
(B24) is the $x$ deformation of (A18) for $m>l$.

\bigskip

2$%
{{}^\circ}%
)\;l>m:$

\bigskip

We define 
\begin{equation}
\Psi _{d,n}\left( y,x\right) =T_{d,n}\left( y,x\right)
+\sum_{i=1}^{l^{\prime }-1}Q^{is}\;W_{i,n^{\prime }}\left( y,x\right) \;T_{d-%
\frac{i}{m^{\prime }},n}\left( x^{\frac{i}{m^{\prime }}}y,x\right)  \tag{B25}
\end{equation}
where $Q=\exp (2i\pi /m^{\prime }).$ The functions $W_{i,n^{\prime }}\left(
y,x\right) $ are constrained by the conditions 
\begin{equation}
\Psi _{d,n}\left( y,x\right) =Q^{s}\;W_{1,n^{\prime }}\left( y,x\right)
\;\Psi _{d-\frac{1}{m^{\prime }},n}\left( x^{\frac{1}{m^{\prime }}}y,x\right)
\tag{B26}
\end{equation}
which\bigskip\ implies 
\begin{equation}
\Psi _{d,n}\left( y,x\right) =\prod_{i=0}^{m^{\prime }-1}W_{1,n^{\prime
}}\left( x^{\frac{i}{m^{\prime }}}y,x\right) \;\Psi _{d-1,n}\left(
xy,x\right)  \tag{B27}
\end{equation}
We obtain the following relations 
\begin{eqnarray}
W_{k,n^{\prime }}\left( y,x\right) &=&W_{1,n^{\prime }}\left( y,x\right)
\;W_{k-1,n^{\prime }}\left( x^{\frac{1}{m^{\prime }}}y,x\right) \;\;  \notag
\\
for\;\;k &=&2,...,m^{\prime }-1,m^{\prime }+1,...,l^{\prime }-1\;\  
\label{B28a}
\end{eqnarray}
\begin{equation}
1+W_{m^{\prime },n^{\prime }}\left( y,x\right) =W_{1,n^{\prime }}\left(
y,x\right) \;W_{m^{\prime }-1,n^{\prime }}\left( x^{\frac{1}{m^{\prime }}%
}y,x\right)  \tag{B28b}
\end{equation}
\begin{equation}
q^{n^{\prime }}\;x^{-\frac{\left( m-1\right) l}{2m}}\;y=W_{1,n^{\prime
}}(y,x)\;W_{l^{\prime }-1,n^{\prime }}\left( x^{\frac{1}{m^{\prime }}%
}y,x\right)  \tag{B28c}
\end{equation}
These three equations are easily transformed into 
\begin{equation}
W_{k,n^{\prime }}\left( y,x\right) =\prod_{i=0}^{k-1}W_{1,n^{\prime }}\left(
x^{\frac{i}{m^{\prime }}}y,x\right) \;\;\;\;k=1,...,m^{\prime }-1  \tag{B29a}
\end{equation}
\begin{equation}
W_{k,n^{\prime }}\left( y,x\right) =\prod_{i=0}^{k-m^{\prime
}-1}W_{1,n^{\prime }}\left( x^{\frac{i}{m^{\prime }}}y,x\right)
\;W_{m^{\prime },n^{\prime }}\left( x^{\frac{k-m^{\prime }}{m^{\prime }}%
}y,x\right) \;\;\;k=m^{\prime }+1,...,l^{\prime }-1  \tag{B29b}
\end{equation}
\begin{equation}
1+W_{m^{\prime },n^{\prime }}\left( y,x\right) =\prod_{i=0}^{m^{\prime
}-1}W_{1,n^{\prime }}\left( x^{\frac{i}{m^{\prime }}}y,x\right)  \tag{B29c}
\end{equation}
\begin{equation}
q^{n^{\prime }}\;x^{-\frac{\left( m-1\right) l}{2m}}\;y=\prod_{i=0}^{l^{%
\prime }-m^{\prime }-1}W_{1,n^{\prime }}\left( x^{\frac{i}{m^{\prime }}%
}y,x\right) \;W_{m^{\prime },n^{\prime }}\left( x^{\frac{l^{\prime
}-m^{\prime }}{m^{\prime }}}y,x\right)  \tag{B29d}
\end{equation}
If we multiply this last equation by ($m^{\prime }-1$) other equations, each
being transformed from the one before by making $y\rightarrow x^{\frac{1}{%
m^{\prime }}}y,$ \ we obtain 
\begin{equation}
x^{-\frac{m^{\prime }-1}{2}}\prod_{i=0}^{m^{\prime }-1}F_{n^{\prime }}\left(
x^{\frac{l^{\prime }-m^{\prime }+i}{m^{\prime }}}y,x\right)
\;\prod_{i=0}^{l^{\prime }-m^{\prime }-1}\left[ 1+F_{n^{\prime }}\left( x^{%
\frac{i}{m^{\prime }}}y,x\right) \right] =q^{n^{\prime }m^{\prime }}\;x^{-%
\frac{\left( m-1\right) l^{\prime }}{2}}\;y^{m^{\prime }}  \tag{B30a}
\end{equation}
where $F_{n^{\prime }}(y,x)=W_{m^{\prime },n^{\prime }}\left( y,x\right) .$
Again, if we multiply equation (B30a) by $m/m^{\prime }$ different
equations, each being transformed from the one before by making $%
y\rightarrow x^{\frac{1}{m}}y,$ \ we obtain 
\begin{equation}
\prod_{i=0}^{m-1}F_{n^{\prime }}\left( x^{\frac{l-m+i}{m}}y,x\right)
\;\prod_{i=0}^{l-m-1}\left[ 1+F_{n^{\prime }}\left( x^{\frac{i}{m}%
}y,x\right) \right] =x^{-\frac{\left( m-1\right) \left( l-1\right) }{2}%
}\;y^{m}  \tag{B30b}
\end{equation}
which is the $x$-deformed equation (A15b). At $\ x=1,\;$equation (B30a) is
polynomial in $F_{n^{\prime }}$ of degree $l^{\prime }.$ For a given $%
n^{\prime },$ we have $l^{\prime }$ solutions; this analysis remains true by 
$x$ deformation and for $x$ not too far from $1.$ Considering the $l^{\prime
}$ solutions $F_{n^{\prime }}^{\left( j\right) },\;\;j=1,...,l^{\prime },$
the system (B25) is a linear system of $l^{\prime }$ equations for $%
l^{\prime }$ unknown functions $T_{d-\frac{i}{m^{\prime }},n}\left( x^{\frac{%
i}{m^{\prime }}}y,x\right) ,\;\;i=0,...,l^{\prime }-1.$ We may write 
\begin{equation}
T_{d-\frac{i}{m^{\prime }},n}\left( x^{\frac{i}{m^{\prime }}}y,x\right)
=Q^{-is}\sum_{j=1}^{l^{\prime }}\Delta _{i,\left( j\right) }^{n^{\prime
}}\left( y,x\right) \;\Psi _{d,n}^{\left( j\right) }\left( y,x\right)
\;\;\;i=0,...,l^{\prime }-1  \tag{B31}
\end{equation}
where $\Delta _{i,\left( j\right) }^{n^{\prime }}\left( y,x\right) \;$is the
ratio of two determinants containing only $W_{i,n^{\prime }}^{\left(
j\right) }\left( y,x\right) $ functions, the powers of $Q$ being extracted
separetely. It is easy to convince oneself that 
\begin{equation}
\Delta _{i,\left( j\right) }^{n^{\prime }}\left( x^{\frac{1}{m^{\prime }}%
}y,x\right) =W_{1,n^{\prime }}^{\left( j\right) }\left( y,x\right) \;\Delta
_{i+1,\left( j\right) }^{n^{\prime }}\left( y,x\right)
\;\;\;\;\;i=0,...,l^{\prime }-2  \tag{B32}
\end{equation}
and that we have the following orthogonality relations 
\begin{eqnarray}
\sum_{i=0}^{l^{\prime }-1}W_{i,n^{\prime }}^{\left( k\right) }\left(
y,x\right) \;\Delta _{i,\left( j\right) }^{n^{\prime }}\left( y,x\right)
&=&\delta _{j}^{k}  \notag \\
\sum_{j=1}^{l^{\prime }}W_{i,n^{\prime }}^{\left( j\right) }\left(
y,x\right) \;\Delta _{k,\left( j\right) }^{n^{\prime }}\left( y,x\right)
&=&\delta _{i,k}  \label{B33}
\end{eqnarray}
Finally, using (B27) and (B29c), we have 
\begin{equation}
\Psi _{d,n}^{\left( j\right) }\left( y,x\right) =\prod_{i=0}^{d-1}\left[
1+F_{n^{\prime }}^{\left( j\right) }\left( x^{i}y,x\right) \right] \;\Psi
_{0,n}^{\left( j\right) }\left( x^{d}y,x\right)  \tag{B34}
\end{equation}
so that the solution of the recursion (B6) is 
\begin{equation}
T_{d-\frac{i}{m^{\prime }},n}\left( x^{\frac{i}{m^{\prime }}}y,x\right)
=Q^{-is}\sum_{j=1}^{l^{\prime }}\Delta _{i,\left( j\right) }^{n^{\prime
}}\left( y,x\right) \;\prod_{k=0}^{d-1}\left[ 1+F_{n^{\prime }}^{\left(
j\right) }\left( x^{k}y,x\right) \right] \;\Psi _{0,n}^{\left( j\right)
}\left( x^{d}y,x\right)  \tag{B35}
\end{equation}
for $i=0,...,l^{\prime }-1.\;$

We must now perform the inverse generalized Fourier transform (B5). Again,
we write $\left( r-1\right) l^{\prime }=Km^{\prime }-h$ where $K$ is an
integer and $0\leq h<m^{\prime }.$ Then equation (B20), where $j$ is summed
from $1$ to $l^{\prime },$ and equation (B21) are valid.\ Again, the initial
conditions that is the values of $Z_{d,r}$\bigskip $\left( y,x\right) $ from
(B1) for $0\leq d<l+1$ determine the functions 
\begin{equation}
\Phi _{n^{\prime }}^{\left( j\right) }\left( y,x\right)
=\prod_{i=0}^{l^{\prime }-m^{\prime }-1}W_{1,n^{\prime }}^{\left( j\right)
}\left( x^{\frac{i}{m^{\prime }}}y,x\right)  \tag{B36}
\end{equation}
\bigskip Using now (B29c), the definition of $F_{n^{\prime }}^{\left(
j\right) }\left( y,x\right) $, (B32) and (B36), we may write 
\begin{eqnarray}
Z_{d,r}\left( y,x\right) &=&\frac{x\frac{r\left( m-r\right) l}{2m}}{m}%
\sum_{n^{\prime },j}q^{-n^{\prime }r}\Delta _{0,\left( j\right) }^{n^{\prime
}}\left( y,x\right) \prod_{k=0}^{d+r-2}\left[ 1+F_{n^{\prime }}^{\left(
j\right) }\left( x^{k}y,x\right) \right]  \notag \\
&&.\prod_{i=0}^{r\left( l^{\prime }-m^{\prime }\right) -1}W_{1,n^{\prime
}}^{\left( j\right) }\left( x^{d+r-1+\frac{i}{m^{\prime }}}y,x\right) 
\label{B37}
\end{eqnarray}
Finally, using (B29d) $r$ times, we get 
\begin{eqnarray}
Z_{d,r}\left( y,x\right) &=&\frac{x^{rd+\frac{r\left( r-1\right) }{2}}y^{r}}{%
m}\sum_{u=0}^{l-1}\Delta _{u}\left( y,x\right) \prod_{k=0}^{d+r-2}\left[
1+F_{u}\left( x^{k}y,x\right) \right] .  \notag \\
&&.\prod_{i=0}^{r-1}\left[ F_{u}\left( x^{d+i-1+\left( r-i\right) \frac{l}{m}%
}y,x\right) \right] ^{-1}  \label{B38}
\end{eqnarray}
where $u=n^{\prime }+\left( j-1\right) \frac{l}{l^{\prime }}.$ Equation
(B38) is the $x$ deformation of (A18) for $l>m$.

\bigskip

\bigskip

\bigskip

\bigskip

\bigskip

\bigskip

$\bigskip $

$\bigskip $

$\left[ 1\right] $ \ \ J.\ M.\ Leinaas and J.\ Myrheim, Nuovo Cimento, 
\textbf{B37}, 1 (1977).

\ \ \ \ \ \ G.\ A.\ Goldin, R.\ Menikoff and D.\ H.\ Sharp, J.\ Math.\ Phys. 
\textbf{22}, 1664

\ \ \ \ \ \ (1981).

\ \ \ \ \ \ F.\ Wilczek,\ Phys.\ Rev.\ Lett.\ \textbf{48}, 1144 (1982); 
\textbf{49}, 957 (1982).

$\left[ 2\right] $ \ \ F.\ Calogero, J.\ Math.\ Phys. \textbf{10}, 2191,
2197 (1969).

\ \ \ \ \ \ B.\ Sutherland, J.\ Math.\ Phys. \textbf{12}, 251 (1971); Phys.\
Rev. \textbf{A4}, 2019

\ \ \ \ \ \ (1971); \textbf{A5}, 1372 (1972).\ 

$\left[ 3\right] $ \ \ F.\ D.\ M.\ Haldane, Phys.\ Rev.\ Lett. \textbf{67},
937 (1991).

$\left[ 4\right] $ \ \ M.\ C.\ Berg\`{e}re, fractional statistic,
cond-mat/9904227.

$\left[ 5\right] $ \ \ C.\ N.\ Yang and C.\ P.\ Yang, J.\ Math.\ Phys.\ 
\textbf{10}, 1115 (1969).

\ \ \ \ \ \ Y.\ S.\ Wu, Phys.\ Rev.\ Lett. \textbf{73}, 922 (1994).

\ \ \ \ \ \ D.\ Bernard, Les Houches, Session LXII (1994).

\ \ \ \ \ \ A.\ Dasni\`{e}re de Veigy and S. Ouvry, Phys.\ Rev.\ Lett.\ 
\textbf{72}, 600 (1994).

$\left[ 6\right] $ \ \ B.\ C.\ Berndt, R.\ J.\ Emery and B.\ M.\ Wilson,
Adv.\ Math.\ \textbf{49}, 123 (1983).

\ \ \ \ \ \ M.\ V.\ N.\ Murthy and R.\ Shankar, exclusion statistics: a
resolution of the

$\;\;\;\;\;\ $problem of negative weights, cond-mat/9903278.

$\left[ 7\right] $\ \ M.\ C.\ Berg\`{e}re, ''composite particles'' and the
eigenstates of

$\;\;\ \ \;\;$Calogero-Sutherland and Ruijsenaars-Schneider,
cond-mat/9907411.

$\left[ 8\right] $\ \ S.\ N.\ M.\ Ruijsenaars and H.\ Schneider, Ann.\
Phys.\ \textbf{170}, 370 (1986).

$\;\;\;\ \ \;$S.\ N.\ M.\ Ruijsenaars, Com. Mat.\ Phys. \textbf{110}, 191
(1987).

$\left[ 9\right] $ \ \ I. G. Macdonald, Symmetric functions and Hall
polynomials, 2nd ed.

$\;\;\;\;\ \;$Clarendon Press (1995).

$\left[ 10\right] \;$W.\ Nahm, A.\ Recknagel and M.\ Terhoeven, Mod.\ Phys.\
Lett. \textbf{A8}, 1835

$\;\;\;\;\;\ $(1993).

\ \ \ \ \ M.\ Terhoeven, Mod.\ Phys.\ Lett. \textbf{A9}, 133 (1994).

\ \ \ \ \ A. G.\ Bytsko, fermionic representation for characters of M(3,t),
M(4,5),

$\;\;\;\;\;\;$M(5,6) and M(6,7) minimal models and related Rogers-Ramanujan
type

\ \ \ \ \ and$\;$dilogarithm identities, hep-th/9904059.

$\left[ 11\right] $ K.\ Hikami, Phys.\ Lett.\ \textbf{A205}, 364 (1995).

$\left[ 12\right] $ R.\ Kedem, T.\ R.\ Klassen, B.\ M.\ McCoy and E.\
Melzer, Phys.\ Lett. \textbf{B304},

$\;\;\;\ \ \ $263 (1993); \textbf{B307}, 68 (1993).

\ \ \ \ \ \ A.\ Berkovich and B.\ M.\ McCoy, the universal\ chiral partition
function

\ \ \ \ \ \ for exclusion statistics, hep-th/9808013.

$\left[ 13\right] $ P.\ Bouwknegt and K.\ Schoutens, exclusion statistics in
conformal field

$\;\;\;\;\ \;\ $theory, hep-th/9810113.

\ \ \ \ \ \ P.\ Bouwknegt, L.\ Chim and D.\ Ridout, exclusion$\ $statistics
in conformal

\ \ \ \ \ \ field theory and the UCPF for WZW models, hep-th/9903176.

$\left[ 14\right] $ M.\ Bauer, C.\ Godr\`{e}che and J.\ M.\ Luck, statistic
of persistent events

\ \ \ \ \ \ in the binomial random walk.\ Will the drunken sailor hit the
sober man?

\ \ \ \ \ \ cond-mat/9905252.

\ \ \ \ \ \ B.\ Derrida and J.\ L.\ Lebowitz, Phys.\ Rev.\ Lett, \textbf{80}%
, 209 (1998).

\newpage
\input epsf.tex
\newcount\figno
\figno=1
\def\fig#1#2#3{
\par\begingroup\parindent=0pt\leftskip=1cm\rightskip=1cm\parindent=0pt
\baselineskip=11pt
\epsfxsize=#3
\centerline{\epsfbox{#2}}
\vskip 12pt
\centerline{{\bf Fig. \the\figno:} #1}\par
\endgroup\par
}
\def\figlabel#1{\xdef#1{\the\figno}}
\def\encadremath#1{\vbox{\hrule\hbox{\vrule\kern8pt\vbox{\kern8pt
\hbox{$\displaystyle #1$}\kern8pt}
\kern8pt\vrule}\hrule}}

\fig{
``Composite particles''
}{
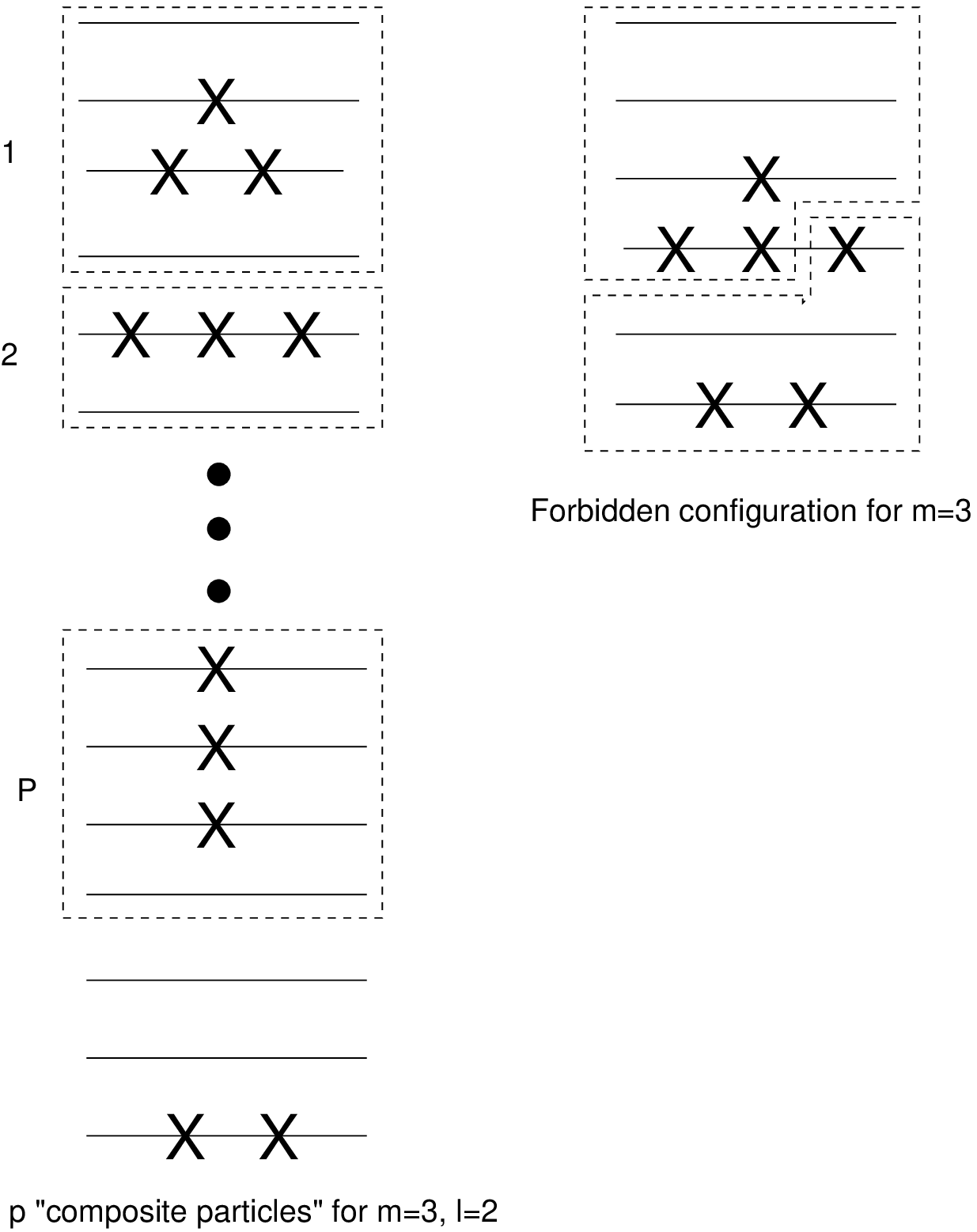
}{
13.116cm
}
\end{document}